\documentclass[twocolumn, 10pt, cleanfoot]{asme2ej}

\usepackage{amsmath,amssymb,amsfonts,bm} 
\usepackage[numbers,sort&compress]{natbib} 
\usepackage{url,doi} 
\usepackage{hyperref}
\usepackage{comment}
\usepackage{gensymb}

\usepackage{subfiles}
\usepackage{graphicx}
\usepackage{caption}    
\usepackage{subfigure} 
\usepackage{float}
\usepackage{physics}
\usepackage{derivative}
\usepackage{bm}
\usepackage[capitalize]{cleveref}
\usepackage[version=4]{mhchem}
 \hypersetup{
    colorlinks = true,
    allcolors = blue
}

\usepackage[dvipsnames]{xcolor}

\renewenvironment{nomenclature}[1][1cm]{%
    
    \section*{NOMENCLATURE}
    \list{}{\leftmargin #1}%
  }%
  {\endlist\par\addvspace{12pt}}

\usepackage{nomencl}
\usepackage{ifthen}
\renewcommand{\nomgroup}[1]{%
\ifthenelse{\equal{#1}{R}}{\item[\textbf{Roman Symbols}]}{%
\ifthenelse{\equal{#1}{G}}{\item[\textbf{Greek Symbols}]}{%
\ifthenelse{\equal{#1}{S}}{\item[\textbf{Subscripts}]}{}}}
}
\makenomenclature

\title{Combustion and Evaporation of Deformable Fuel Droplets}

\author{Meha Setiya\thanks{Address all correspondence to this author.}
    \affiliation{
	PhD Candidate\\
	Department of Mechanical Engineering\\
	Virginia Tech \\
	Blacksburg, Virginia, USA 24061\\
    Email: setiyameha@vt.edu
    }	
}

\author{John Palmore Jr
    \affiliation{Assistant Professor\\
    Department of Mechanical Engineering\\
	Virginia Tech \\
	Blacksburg, Virginia, USA 24061\\
    Email: palmore@vt.edu
    }
}

\begin{document}

\maketitle
\begin{abstract}
\textit{This study focuses on combustion and evaporation of an isolated freely deforming fuel droplet under convective flow. The droplet shape is modified by varying Weber number at moderate Reynolds numbers. A simplified chemical reaction mechanism is used for combustion modelling.}
\textit{The Direct Numerical Simulation (DNS) results show a net positive effect of Weber number on total evaporation rate ($\dot{m}$) for both pure evaporation and combustion cases. The enhancement in $\dot{m}$ for higher Weber number reaches upto $9 \%$ for combustion. A non-spherical envelope flame is observed which grows with time. The Damk\"{o}hler number is higher than 1 for this flame type which leads to faster reaction rates in comparison to evaporation. Hence, the combustion process is seen to be unaffected by droplet shape. 
An additional comparison between 3-D and 2-D  combustion results is performed to understand if 2-D studies can reflect the right physical aspects of this problem. It is found that local evaporation flux in 2-D is $ 42.5 \%$ lower due to lower temperature gradients near the droplet surface for the same inflow velocity. The deformation of droplet is significantly different in 2-D which affects the boundary layer development and the wake flow. This is seen to affect the flame shape at the downstream of droplet. Hence, the 2-D simulations do not recover the correct behaviors.}

\end{abstract}

\nomenclature[Rp]{$Y_i$}{Specie mass fraction of species `i'}
\nomenclature[Rp]{$W_i$}{Molecular weight of species  `i'}
\nomenclature[Rp]{$d_0$}{Initial diameter of the droplet $[m]$}
\nomenclature[Rp]{$U$}{Incoming air velocity $[m/s]$}
\nomenclature[Rp]{$P$}{Pressure $[Pa]$}
\nomenclature[Rp]{$T$}{Temperature $[K]$}
\nomenclature[Rp]{$k$}{Thermal conductivity $[W/(m\cdot K)]$}
\nomenclature[Rp]{$A$}{Pre-exponential factor $[(mol/m^3)^{0.75} (1/s)]$}
\nomenclature[Rp]{$c_P$}{Specific heat $[J/(kg\cdot K)]$}
\nomenclature[Rp]{$E_a$}{Activation energy $[J/mol]$}
\nomenclature[Rp]{$R_u$}{Universal gas constant $[J/(mol.K)]$}
\nomenclature[Rp]{$B_q$}{Spalding heat transfer number}
\nomenclature[Rp]{$h_{fg}$}{Enthalpy of vaporization $[J/kg]$}
\nomenclature[Rp]{$\Delta{h_c}$}{Lower heating value of fuel $[kJ/kg]$}
\nomenclature[Rp]{$We_0$}{Initial Weber number}
\nomenclature[Rp]{$Re_0$}{Initial Reynolds number}
\nomenclature[Rp]{$Da_0$}{Initial Damk\"{o}hler number}
\nomenclature[G]{$\alpha$}{Liquid volume fraction}
\nomenclature[G]{$\kappa$}{Local curvature}
\nomenclature[G]{$\sigma$}{Surface tension $[N/m]$}
\nomenclature[G]{$\rho$}{Density $[kg/m^3]$}
\nomenclature[G]{$\mu$}{Viscosity $[kg/(m\cdot s)]$}
\nomenclature[G]{$\nu$}{Air to Fuel ratio}
\nomenclature[G]{$\phi$}{Equivalence ratio}
\nomenclature[G]{$\dot{\omega}$}{Chemical reaction rate $[mol/m^3 \cdot s]$}
\nomenclature[S]{$i$}{Specie number}
\nomenclature[S]{$L$}{Liquid}
\nomenclature[S]{$G$}{Gas}
\nomenclature[S]{$\infty$}{Far field condition}
\printnomenclature

\section{Introduction} \label{sec:intro}
In case of combustion in gas turbine, a liquid fuel jet is injected into the hot stream of incoming air. This leads to a series of complex physical processes, including primary atomization and  secondary atomization. During the secondary atomization, the liquid fuel ligaments further break up into smaller droplets due to imbalance of disruptive inertial force and surface tension \cite{combustion_lefebvre}. The droplet size in this regime varies from few tens of micrometers to few hundreds of micrometers \cite{sirignano_1999}. Although the larger droplets are fewer in the count, they contribute to the major portion of the spray mass as this will scale with the cube of droplet diameter. For example, the volumetric contribution of a single droplet of $d=1000 \ \mu m$ is equivalent to $10^6$ times of small droplets of size $d=10 \ \mu m$. Because of this fact, the diameter of the mean droplet by mass is significantly larger than the mean particle diameter (as seen in \cite{book_lefebvre}). As these spray droplets are the rate limiter for combustion \cite{Faeth2002}, in order to improve the overall understanding, the study of these large droplets is important.

Due to the turbulent and convective flow conditions inside the combustion chamber, the droplet shape is governed by the aerodynamic stresses and surface tension at the liquid-gas interface \cite{frankel1983}. This can be written in terms of a non-dimensional number, Weber number ($We= \frac{\rho_G {U_\infty}^2 d_0}  {\sigma}$) which is a ratio of these two forces. In case of the small droplets, they will have higher surface tension due to higher curvature and small Weber numbers. Hence, they tend to remain in a nearly spherical shape. However, the larger spray droplets will have a higher Weber number and therefore higher tendency to deform. 

Significant experimental as well as numerical work exists on combustion of ``spherical droplets" \cite{KOTAKE1969595,RAGHAVAN2005,KUMAGAI1957726,Pope2005,wu2010}. However, there is a dearth of literature on involving ``deformable droplets". As the shape of a droplet is one of the parameters can alter the flow development around the droplet and hence can affect its combustion. Therefore, it is important to address this gap. In effort to address this gap, this paper reviews the major past work on evaporation of deformable droplets first, followed by combustion studies. 

Recent analytical studies by Tonini and Cossalli \cite{tonini2013, tonini2016} suggested that the iso-volume spheroidal droplets (prolate and oblate) evaporate faster in comparison to spherical droplet. The local flux was found to be proportional to surface curvature under the stagnant gas environment. Furthermore, the results of analytical study using perturbation theory by Palmore \cite{palmore_JHT_2022} demonstrated the freely deformed droplets has higher evaporation rate in comparison to isovolume spherical droplet. Although, the surface area is higher for deformed droplet, the total evaporation rate of a non-spherical droplet does not scale by the ratio of surface areas times the evaporation rate of spherical droplet. At low Reynolds number (nearly quiescent flow conditions), the non-dimensional total evaporation rate given by Equation 72 in \cite{palmore_JHT_2022} suggests that, it has a complex relationship with the droplet shape (Liquid-Gas interface) governed by Weber number. It was also shown that the deformation of the droplet as represented by Weber number, enhances the combustion rate by increasing the rate of fuel evaporation. This alters the distance of the flamesheet from the droplet surface and hence, it changes the entrainment of ambient gas between flamesheet and droplet surface. However, the results from his work were limited to combustion in stagnant air ($Re=0$). 
One of our recent work on pure evaporation of freely deforming droplets under convective flow in \cite{SETIYA2023104455} demonstrated that surface averaged total evaporation rate is higher for the deformed droplets than the spherical droplets. This enhancement is associated with the local curvature in the front region and wake flow interaction with the droplet. Both these parameters are related to droplet shape.

In case of combustion in a stagnant flow, many experimental studies such as work by Xia \cite{xia2015computational} and Pacheo et al. \cite{Pacheco_2021} focused on analyzing the dynamics and chemistry of spherical diffusion flames. 
A relatively larger droplet size was chosen in the range of $100 \ \mu m$ to $10\ mm$ for the majority of the experimental work as they generally try to mimic the initial Reynolds numbers ($Re_0$) as seen by the real spray droplets. However, matching $Re_0$ with the larger droplets in experiments mean slower inflow velocities, hence lower Weber numbers. Therefore, the droplet shape is nearly spherical.

A few experimental studies on the burning of fuel particles such as \cite{SATO1990142, Gokalp1989,GOLLAHALLI1974313, GOLLAHALLI1975409, RAGHAVAN2005} covered the burning under a convective flow. The effect of natural convection on high pressure droplet combustion was studied by Sato et al. \cite{SATO1990142}. For pressure below critical pressure of liquid fuel, the burning time decreases and above critical pressure, burning time increases. The presence of natural convection affects the heat and mass transfer across the flame. 

Gokalp et al. \cite{Gokalp1989} conducted low temperature droplet burning under normal and reduced gravity conditions under natural and forced convection. They found that the assumption of spherical flame no longer holds for convective conditions. 
Moreover, due to presence of silica filament to hold the droplet, the droplet shape was closer to an ellipsoid than a sphere in their experiment. The burning rates or surface regression rates using $d^2$ law involved calculation of characteristic dimension based on equivalent surface area or volume of a sphere. However, the effect of droplet shape specifically on burning rate was not the focus of their work.

The studies by Gollahalli and Brzustowski \cite{GOLLAHALLI1975409} suggested that the flow conditions drive the nature of the diffusion flame, either an envelope flame or a wake flame or transitioning from envelope flame to wake flame. Such flame development is sensitive to thermophysical properties of ambient gas as well as its inflow velocity. If the inflow velocity is lower than a critical value, envelope flame will be seen. If the gas velocity is higher than a critical value, the wake flame will be observed. Moreover, a sudden drop in evaporation rate was observed when transitioning from envelope flame to wake flame.

Raghvan et al. \cite{RAGHAVAN2005} studied the effect of ambient temperature on the burning rates for different inflow velocities and various flame shapes were observed. On the similar lines, a numerical study by Wu and Sirignano \cite{wu2010} on transient burning of n-octane was performed at high temperature and $20 \ atm$ pressure air stream. Specifically, the transient shape of flame, surface temperature and burning rates were studied at different initial Reynolds numbers $Re_0 = \rho_\infty U_{rel} d_0/ \mu_\infty$ and initial Damk\"{o}hler numbers (${Da}_0$) for $d_0=50 \ \mu m$. The initial Damk\"{o}hler number is defined as the ratio of reaction rate and convective transport rate. 
\begin{equation}
    {Da}_0 = \frac{d_0/U_\infty}{\rho_\infty Y_F^0/\dot{\omega_0}W_F}
\end{equation}
The results showed that, for given flow conditions, if an envelope flame is observed, it is going to stay an envelope flame through the lifetime of the droplet. However a wake flame can either transition to an envelope flame at the later of lifetime of droplet or can extinguish. This numerical study suggested a critical initial ${Da}_0 = 1.02$ for n-octane at $P_\infty=20\ atm$ and $T_\infty=1500 K$.  An envelope flame is observed above at ${Re}_0 = 11$, $Da_0 =1.2$ and wake flame is seen at ${Re}_0=45$, $Da_0=0.3$. Another numerical study by Pope and Gogos \cite{Pope2005} analyzed the extinction of the envelope flame for various sizes of n-heptane droplet ranging from $d_0= 0.1 \ mm \ - \ 3 \ mm$ at atmospheric pressure and varying temperature conditions. For smaller droplets of $d_0 < 0.1\ mm$, the extinction velocity was found to be proportional to $d^{0.5}$ and the correlation involving $Da_0$, $Re$ and transfer number $B$ was developed. It is important to note that all these numerical studies were performed with the assumption of spherical (non-deforming) droplet shape under the axisymmetric flow.

Based on this literature review, the interaction of the combustion of deformable droplet under a convective flow environment is not yet unknown. Hence, this work covers the impact of droplet shape on its pure evaporation and combustion under convective flow of moderate Reynolds number using DNS. The droplet shape is varied by modifying the Weber number.
This work is organized as follows: \cref{sec:problem} covers the problem formulation and summarizes the flow solver details, \cref{sec:num_setup} describes the numerical setup including the boundary conditions and the thermo-physical conditions of fluids, this is followed by results and discussion in \cref{sec:results} and conclusions in \cref{sec:conclusion}.


\section{Problem Formulation} \label{sec:problem}
In the current problem, the combustion of an isolated fuel droplet of n-decane is studied. The initial droplet diameter is $d_0= 100 \ \mu m$. The flow over the droplet is mimicked which is falling at its terminal velocity ($U_T$).
This work uses a numerical framework developed in \cite{palmore_2019} for the analysis of the given problem. Brief details of the same is mentioned in \ref{section:flowsover}.
Subscripts $L$, $G$ refer to liquid and gas phase respectively throughout this work.

\subsection{Flow Solver} \label{section:flowsover}
A finite volume based framework used in this work. This is built upon NGA, an in-house code for simulating low-Mach number multiphase flows. It uses interface-capturing Direct Numerical Simulation (DNS) in which dynamics of flow are solved using first principles i.e. conservation of mass, momentum, and energy. These can be described in their mathematical form as follows. 

The conservation of momentum for a Newtonian fluid can be written as,
\begin{subequations}
\begin{align}
  \frac{\partial\left(\rho\bm u\right)}{\partial t}+\nabla\cdot\left(\rho\bm u\otimes\bm u\right)=-\nabla p+\nabla\cdot\left(\mu\bm S\right), \text{where}
  \label{eq:NS1}
  \\
  \bm S=\left(\nabla\bm u+{\nabla\bm u}^\top-\frac{2}{3}\nabla\cdot\bm u\right)
  \label{eq:NS2}
\end{align}
\end{subequations}
Here, $\rho$ and $\mu$ are the fluid density and dynamic viscosity; $\bm u$ is the velocity, and $p$ is the pressure.

In order to conserve the momentum, Equation (\ref{eq:NS1}) must be coupled with the continuity equation. For incompressible flow, continuity equation can be written as,
\begin{align}
  \nabla\cdot \bm u =0 \label{eq:cont}
\end{align}
The pressure is computed from the incompressibility constraint for the velocity field and is solved using Poisson Equation \cite{palmore_2019}. The scalar quantities like temperature and chemical species are solved using two field approach \cite{MA2013552}.

For low-Mach number flow, the conservation of energy in the liquid phase can be written as,
\begin{align}
\frac{\partial (\rho_L\ c_{p,L}\ T_L)}{\partial t} +\nabla\cdot(\rho_L\ c_{p,L}\ T_L \ \bm u) =  \nabla \cdot(\rho_L\ c_{p,L}\lambda_L \nabla T_L) \label{eq:liq_energy}
\end{align}

Here $\lambda_L = \ k_L / (\rho_L \ c_{p,L})$, known as thermal diffusivity. 

The energy equation for gas phase can be written in the same way.
\begin{align}
\frac{\partial (\rho_G\ c_{p,G}\ T_G)}{\partial t} +\nabla\cdot(\rho_G\ c_{p,G}\ T_G \ \bm u) =  \nabla \cdot(\rho_G\ c_{p,G}\ \lambda_G \nabla T_G)  + {\dot{\omega}_T}
\label{eq:gT}
\end{align}
Here ${\dot{\omega}_T}$ is the heat source term due to combustion.

The equation for chemical species in the gas phase can be specified as,
\begin{align}
\frac{\partial (\rho Y_i) }{\partial t} +\nabla \cdot(\rho  Y_i \bm u) =  \nabla .(\rho D \ \nabla Y_i) +{\dot{\omega}_i} \label{eq:specie}
\end{align}
Here $Y_i$ is $``i"$ species mass fraction and ${\dot{\omega}_i}$ is the chemical source term for each species and $D$ is diffusion coefficient. These source terms are explained in \cref{sec:combustionmech}.

In case of receding L-G interface, for constant thermophysical properties, the conservation of liquid volume fraction ($\alpha$) is given by
\begin{align}
\frac{\partial \alpha }{\partial t} +\nabla \cdot(\alpha \bm u) = - \frac{\dot{m''}}{\rho_L} \delta
\label{eq:alpha}
\end{align}
Here, the surface area density $\delta$ for any arbitrary computational cell $\Omega$ is defined as,
\begin{align}
     \delta= \int_{\Gamma \ \cap \ \Omega} dS \bigg / \int_\Omega  dV \label{eq:delta} 
\end{align}
And, $\dot{m''}$ is local vaporization rate per unit area which satisfies,
\begin{align}
     \odv {}{t} \int_L \rho_L dV = - \int_\Gamma \dot{m''} dS \label{eq:mdot} 
\end{align}
 
Here the subscripts $\Gamma$ is referred to variables measured at interface. 
Equation  \ref{eq:gT} and \ref{eq:specie}, must satisfy the interface conditions,
\begin{align}
  \dot{m''} = \frac{[\bm n_\Gamma \cdot \rho_G \ D \nabla Y_F]_\Gamma}{Y_{F,\Gamma}-1} = \frac{[\bm n_\Gamma \cdot k  \nabla T]_\Gamma}{h_{fg}} 
\end{align}
To ensure the matching conditions at liquid-gas interface, the jump condition for mass can be written as,
\begin{align}
[\bm n.\bm u]_\Gamma = \dot {m''} \bigg[\frac{1}{\rho}\bigg]_\Gamma
\end{align}
Here the bracket notation is used for the jump across the interface specified as $[\phi]_\Gamma= \phi_G-\phi_L$. \\
The jump condition for momentum is
\begin{equation}
[\ p]_\Gamma = -\sigma \kappa - \dot{m''^2}\bigg[\frac{1}{\rho}\bigg]_\Gamma
\label{eq:PJ}
\end{equation}
where $\sigma$ is surface tension and $\kappa $ is local curvature of the interface and is positive when the liquid is locally convex. The details of implementation of Ghost Fluid Method for matching conditions is given in \cite{palmore_2019}.

\subsection{Combustion Mechanism} \label{sec:combustionmech}
A single-step mechanism \cite{westbrook1981} is used for the combustion of the droplet. Five chemical species in total are solved in gas phase, namely fuel (\ce{F}), oxygen (\ce{O2}) as oxidizer, nitrogen (\ce{N2}) and products of combustion, water (\ce{H2O}), and Carbon dioxide (\ce{CO2}).
\begin{equation}
    \ce{C10H22} +15.5 \ \ce{O2} \longrightarrow 10 \ \ce{CO2} + 11 \ \ce{H2O}
\end{equation}
The sum of all species mass fraction sums to unity. 
\begin{equation}
    \sum_{i} Y_i = 1
\end{equation}
Nitrogen is an inert gas and does not participate in combustion. Hence, the species mass fraction of Nitrogen ($Y_{N_2}$) will be computed as:
\begin{equation}
    Y_{\ce{N2}} = 1-(Y_F+Y_{\ce{O2}}+Y_{\ce{CO2}}+Y_{\ce{H2O}})
\end{equation}
Therefore, only 4 species need to be solved inside the gas phase.

Burning of the fuel produces heat due to combustion and its value is $\Delta h_c= 44602 \ [kJ/kg] = \ 6305300.0 \ [J/mol]$ for n-decane. The fuel consumption rate  ($\dot\omega_F$) depends on the temperature and pressure of ambient oxidizer and the species mass fraction of fuel ($Y_F$) and oxidizer ($Y_{O_2}$) and is defined as:
\begin{equation} 
    \dot{\omega}_F = W_F \times (- A \ e^{-E_a/R_u \ T_\infty} [\rho^* \ Y_F / W_F]^{0.25} [{\rho}^* \ Y_{\ce{O2}} / W_{\ce{O2}}]^{1.5}) 
    \label{eq:1-step}
\end{equation}
Here, negative sign associated with $\dot{\omega_F}$ is due to consumption of the fuel. The molecular weight of fuel is $W_F=  0.0142 \ [kg/mol]$  and the unit for $\dot{\omega}_F$ is [$kg/(m^3 \cdot s)$]. The pre-exponent factor $A=12016655 \ [(mol/m^3)^{1-m-n} 1/s]$, where $m=0.25,\  n=1.5$. The activation energy required for the combustion to take place is $E_a=125520.0 \ [J/mol]$, $R_u = 8.314 \ [J/(mol \cdot K)]$ is universal gas constant. 

The density $\rho^*$ is used in combustion calculations to take into account of mixture properties. It is calculated using ideal gas law and mixture properties as expressed in  \cref{eq:rho*}. 
\begin{equation}
    \rho^* = \frac {P_\infty}{R_u T_\infty (\sum_i Y_i/W_i)}
    \label{eq:rho*}
\end{equation}
It is worth highlighting that $\rho^*$ is different from $\rho_G$ which is a constant reference gas density. 
Similarly to \cref{eq:1-step}, the oxidizer consumption rate, production rate of water and carbon dioxide can be written as:
\begin{subequations}
\begin{equation}
    \dot\omega_{\ce{O2}} = (W_{\ce{O2}}/W_F) \times (15.5 \times \dot{\omega_F})
\end{equation}
\begin{equation}
    \dot\omega_{\ce{H2O}} = (W_{\ce{H2O}}/W_F) \times  (- 11 \times \dot{\omega_F})
\end{equation}
\begin{equation}
    \dot\omega_{\ce{CO2}} = (W_{\ce{CO2}}/W_F) \times (- 10 \times \dot{\omega_F})
\end{equation}
\end{subequations}
Moreover, the temperature production term can be written as
\begin{equation}
    \dot\omega_{T} = -W_F \times \bigg(\frac{\Delta h_c  \ \dot{\omega_F}}{\rho_\infty {c_p}_G}\bigg)
\end{equation}

As the combustion timescale is much faster than the fluid timescales, advancing the flow using combustion timescale would make the solution computationally intractable. Hence, a segregated timestep advancement model is used. At each time step, a stiff integrator is used to integrate the chemical reaction sources assuming the fluid flow is stagnant. Then the integrated sources are coupled into the fluid and scalar transport equations. This strategy is common in the combustion simulation community, hence not explained in detail here. More details on our implementation of the idea can be found in our previous works \cite{palmore_combustion_2019}.

This numerical framework has been validated for pure evaporation under stagnant in \cite{palmore_iclass} and in convective flow \cite{SETIYA2023104455}. 

\section{Numerical Setup} \label{sec:num_setup}
The numerical setup for this problem includes an isolated droplet located at the center of the domain. This study improves on previous numerical studies by not imposing any droplet shape such as sphere or ellipsoid. Instead the droplet freely deforms based on the aerodynamic forces it experiences. The initial droplet diameter is $d_0 = 100 \ \mu m$ and it is located at the center of the domain.

\begin{figure} [h!]
   \centering
   \includegraphics[width=1\linewidth]{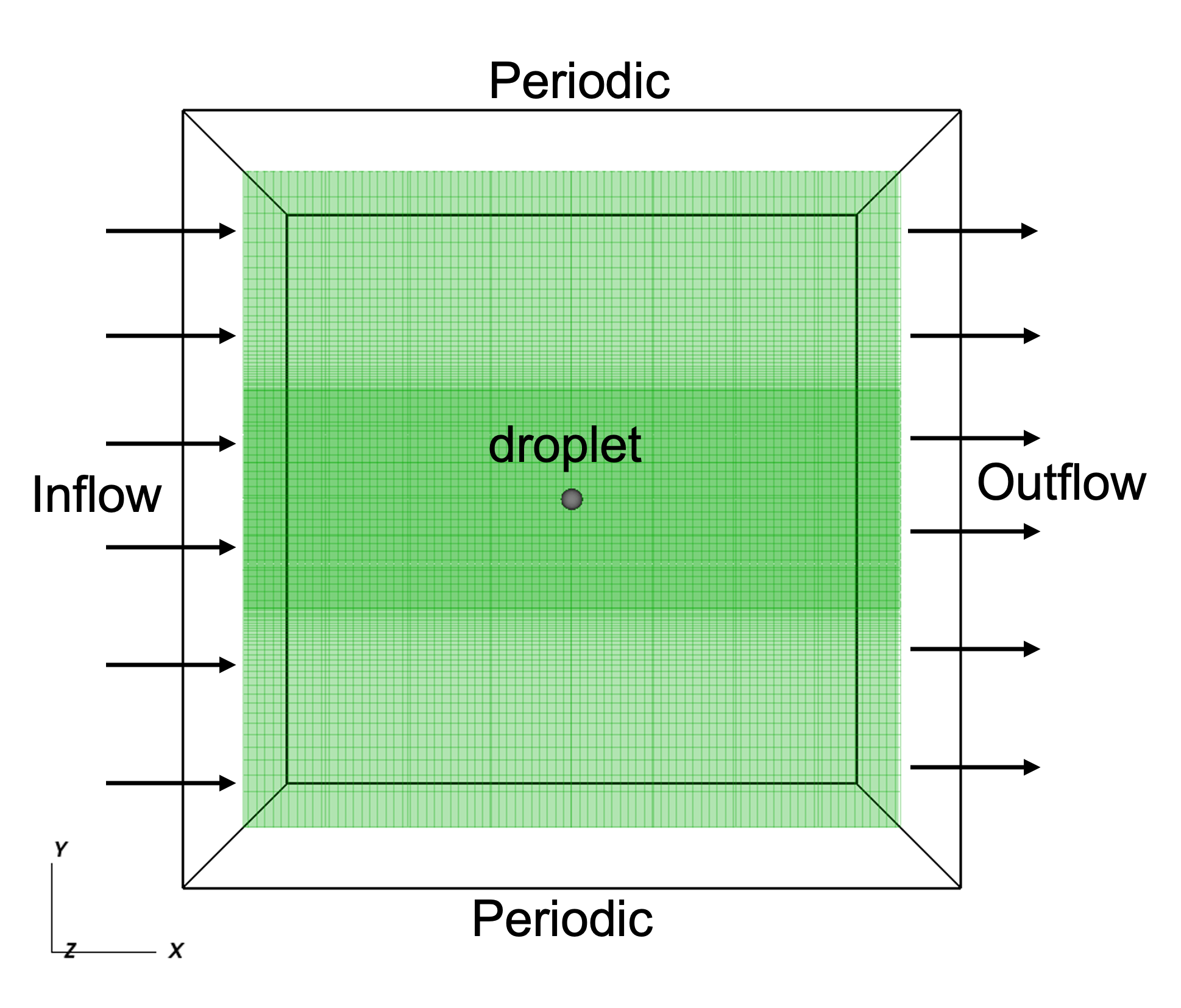} 
   \caption{A sketch of numerical setup and boundaries}
   \label{fig:3d_domain}
\end{figure}

The domain size is $30d_0 \times 30d_0 \times 30d_0$ for these studies. A structured non-uniform grid is used for these studies. The dimension of uniform grid region are $30d_0 \times 10d_0 \times 10d_0$ and beyond this region, a stretch ratio of 1.09 is applied in order to optimize the computational accuracy and expense as shown in \cref{fig:3d_domain}. The grid size in the uniform grid region is  $dx = dy = dz = 5 \times 10^{-6} \ m$. Based on the analytical calculation for vapor film thickness around a non-evaporating sphere in convective flow \cite{abramzon1989}, this grid resolution leads 10 cells to cover the vapor film around the droplet which is sufficient to resolve the thermal and vapor film thickness. It is important to note that the domain size is selected such a way that boundaries do not influence the flow around droplet. 

As shown in \cref{fig:3d_domain}, a fixed inflow boundary condition is applied on the left face of the domain. The right face is used as a convective outflow. The remaining boundaries in y and z directions are treated as periodic. In order to keep the droplet at its predetermined location through out the simulation, ``gravity update" method is used. This method mimics the flow over a freely-deforming droplet falling at its terminal velocity. The artificial gravity on the droplet is applied in feedback loop manner in order to accommodate the droplet center shift in time. This method is seen to be effective in minimizing the droplet shift. More details of this can be found in our previous work \cite{setiya_method_2020, yushu_scitech}.

The ambient conditions for the numerical studies are  $P_\infty = 20 \ atm$ and high temperature $T_\infty= 1400 \ K$ \cite{wu2010}. The droplet is at its saturated conditions. It evaporates and burns in a convective flow with $Y_{\ce{O2}} =0.23$, $Y_{\ce{F}} =0$, $Y_{\ce{N2}}:0.77$, and $T=T_\infty$ at inlet. Reynolds number based on the initial diameter, defined as $Re_0 = \rho_\infty U_{rel} d_0/ \mu_\infty$, here $U_{rel}$ is the relative velocity of oxidizer ; $U_{rel}= U_\infty - U_{drop}$. As the droplet stays in the center of the domain, $U_{drop}$ is negligible and hence, for a constant inflow velocity, $U_{rel}$ does not change significantly. $Re_0=25$ is considered for these studies as the flow over sphere at this $Re_0$ is nearly attached \cite{taneda1956}. To study the effect of deformation, the initial Weber number $We_0= \rho_\infty U_{rel}^2 d_0 /\sigma$ is varied by modifying the surface tension of liquid. The range for Weber number is $We_0=1-12$. The upper limit of $We_0=12$ is imposed due to the droplet breakup beyond this value \cite{loth2006}. The liquid properties refer to fuel properties and are taken at boiling temperature $T_L= 615 \ K $ at $P_\infty=20 \ atm$ \cite{nist_webbook, Yaws1999} and gas properties refer to air and are taken $P_\infty=20 \ atm$ and $T_\infty=1400 \ K$ as mentioned in \cref{table:1}.
\begin{table}
\caption{Fluid properties at $P_\infty=20 \ atm,\ T_{boil}=615\ K,\ T_G=1400\ K$}
\label{table:1}
\begin{center}
\begin{tabular}{ |p{3cm}|p{2cm}|p{2cm}| } 
\hline
Property & Air & Fuel\\ 
\hline
Temperature $(T)$  & 1399.99  & 614.05 \\ 
Density $(\rho)$  & 5.04 & 300.00\\ 
Viscosity $(\mu)$  & $5.05 \times 10^{-5}$ & $4.00 \times 10^{-4}$\\ 
Specific heat $(c_P)$ & 1097.28 & -- \\ 
Thermal conductivity $(k)$  & 0.087 & 0.079\\ 
Latent heat of vaporization  $(h_{fg})$ & -- &48126 \\ 
Adiabatic flame temperature $(T_{ad})$ & -- & 2286 \\
\hline
\end{tabular}
\end{center}
\end{table}
\subsection{Analytical Calculation of Gas Temperature} \label{burning_model}
\begin{figure} [h!]
   \centering
   \includegraphics[width=1\linewidth]{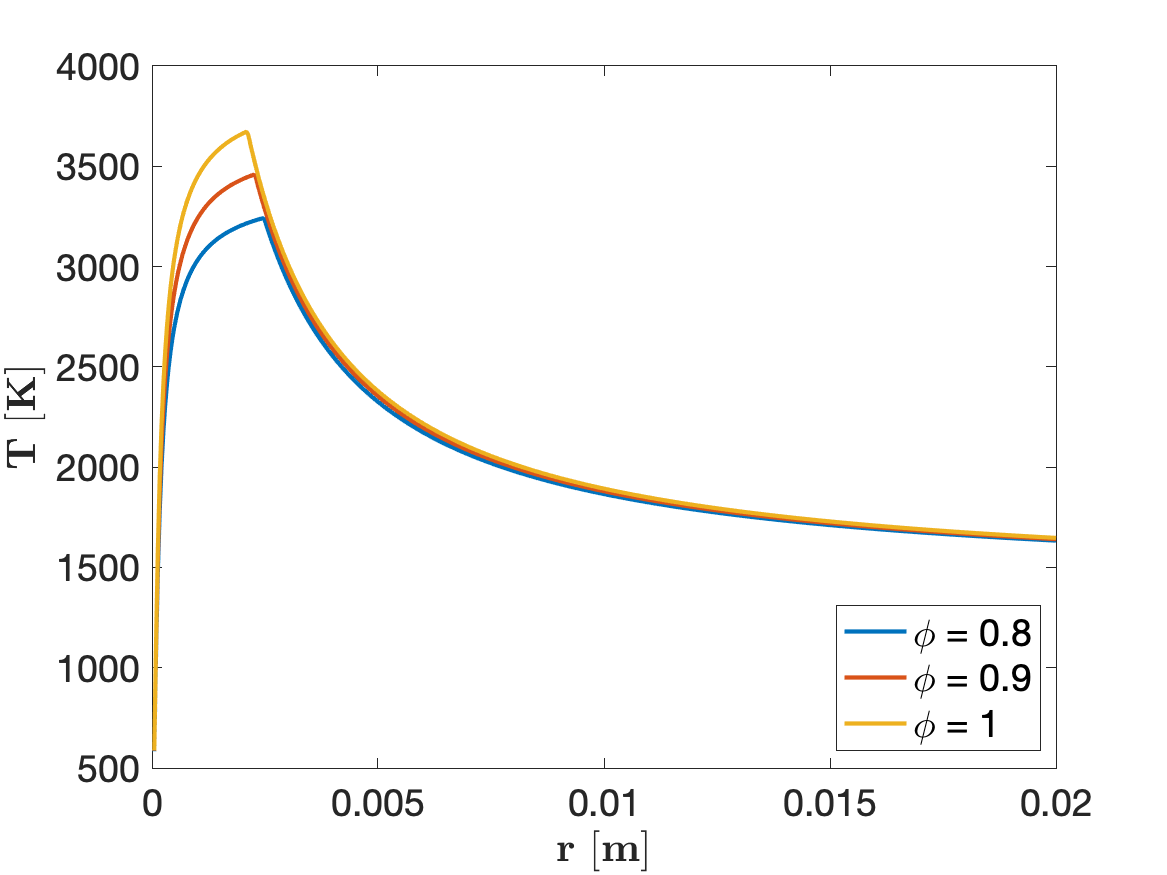} 
   \caption{Simple droplet burning model: Temperature variation in the radial direction \cite{turns2000}}
   \label{fig:simple_burning_model}
\end{figure}

The solver uses the BQUICK algorithm for scalar transport \cite{Herrmann2006}. The algorithm is designed to ensure that transported scalars remain strictly bounded between a given minimum and maximum value. For the species mass fraction, the natural choice for these bounds are 0 and 1. For the temperature, the minimum temperature of $600 \ K$ is selected which is lower than the boiling temperature of the droplet. The highest temperature in this process is the flame temperature due to burning. To evaluate the flame temperature analytically for the given ambient conditions, a simple 1-D droplet burning model is used \cite{turns2000}. The results of the model are demonstrated in  \cref{fig:simple_burning_model}. Moving outward from the droplet surface, the temperature increases rapidly towards a peak value, which can be identified as the flame temperature. The temperature then decreases smoothly with increasing distance from the droplet surface. Various local equivalence ratio exist during the combustion process, hence the flame temperature will be function of the equivalence ratio is defined as $\phi = (A/F)_{st}/ (A/F)$. Here ($A/F$) is the actual air to fuel ratio whereas $(A/F)_{st}$ is the stoichiometric ratio $(A/F)_{st} = 15$. For the flammability limits $\phi_L =0.8$ and $\phi_U=1$ for a diffusion flame \cite{RAGHAVAN2005}, we found that the flame temperature varies from $\sim 3200 - 3660 K$ as shown in \cref{fig:simple_burning_model}. Though it is an oversimplified model, this 1-D calculation helps in estimating the highest temperatures that may be observed in the numerical simulations and use an appropriate initial maximum temperature accordingly. With a great amount of margin, the maximum domain temperature is taken as $5000 \ K$ for these studies. 

\section{Results and Discussion} \label{sec:results}
\subsection{Effect of Droplet Shape on Pure Evaporation and Combustion}
As the combustion process is limited by the evaporation rate of liquid fuel, it is important to study the effect of droplet shape on pure evaporation first under the given ambient conditions. As the droplet shape is governed by its initial Weber number, two cases with $We_0 = 1$ and $We_0=12$ are studied at $Re_0 = 25$ for given ambient conditions. The droplet is initially spherical in shape and eventually deforms freely due to the imbalance of forces.

\begin{figure}[h!]
   \centering
   \includegraphics[width=1\linewidth]{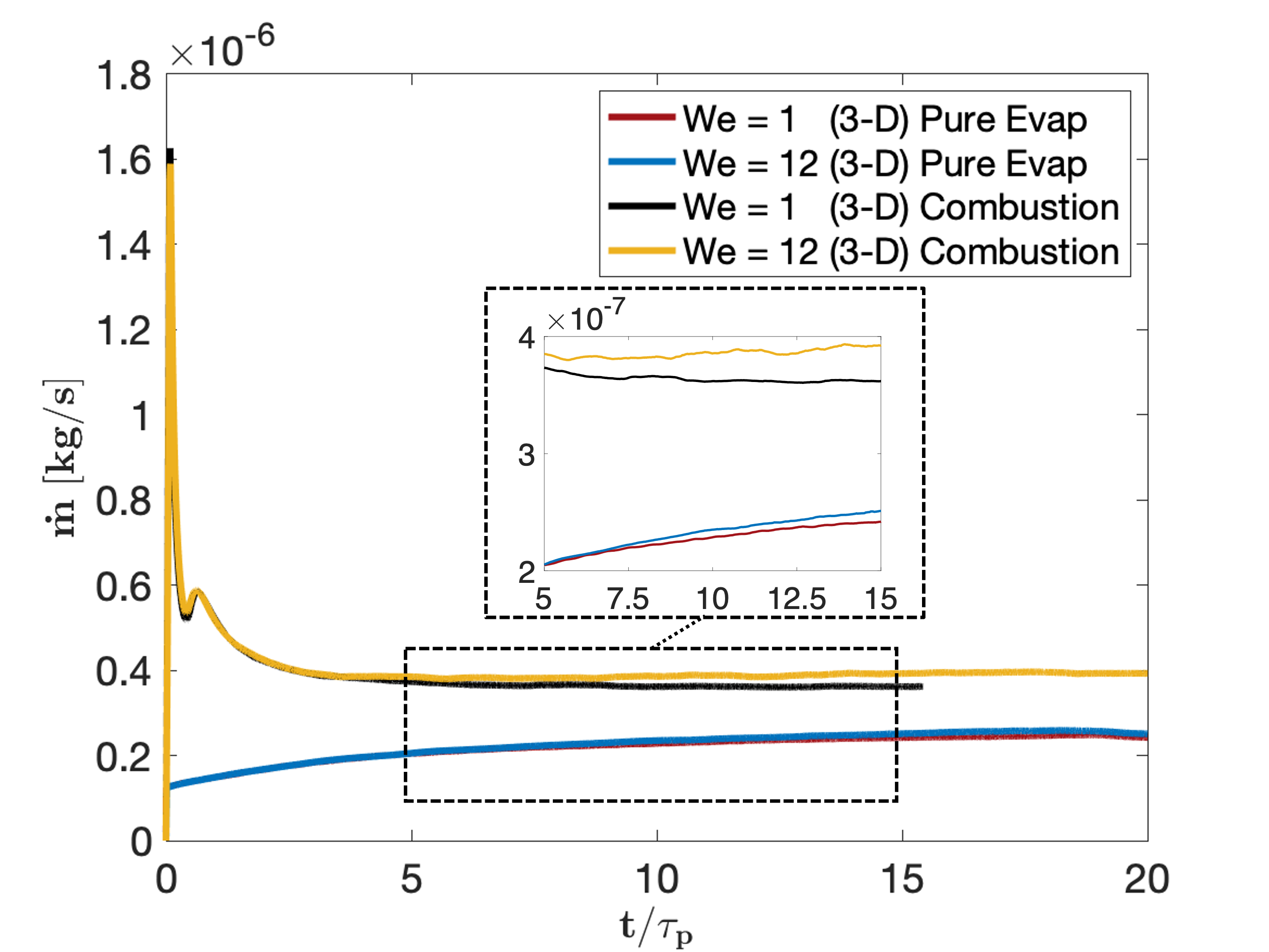} 
   \caption{Total evaporation rate $[kg/s]$ with $t/\tau_p$ at $Re_0=25$ }
   \label{fig:3d_weber_mdot}
\end{figure}

To quantify the effect of droplet shape on pure evaporation and combustion, total evaporation rate ($\dot{m} \ [kg /s]$) is plotted in \cref{fig:3d_weber_mdot}. The results show time history of surface averaged total evaporation rate over time for both pure evaporation and combustion. The time on x-axis is normalized using a capillary response time ($\tau_p$) for droplet which is defined as 
\begin{equation}
\tau_p = \sqrt{\frac{\rho_L +\rho_G}{\sigma}} {\bigg(\frac{d_0}{2\pi}}\bigg)^{3/2} 
\end{equation}
In these results, the instantaneous evaporation rate starts at zero and rapidly increases. Since the droplet is preheated to the saturation temperature, this period is due to the adjustment of the gaseous flow field around the droplet. After a small period of time, $\dot{m}$ reaches a steady value for both pure evaporation and combustion cases. In terms of effect of Weber number on $\dot{m}$ at this $Re_0$, the pure evaporation cases have slightly higher value of $\dot{m}$ for higher Weber number, whereas in case of combustion $We_0=12$ show significantly higher $\dot{m}$. An increase of $4.85 \%$ is seen at time instant $t/\tau_p \sim 7$. This enhancement increase with time and reaches upto $\sim 9 \%$ at $t/\tau_p \sim 15$ as shown in the zoomed-in plot. Moreover, $We_0=1$ combustion case have $11.8\%$ higher $\dot{m}$ when compared with pure evaporation. This is due to overall higher temperature of ambient gas around the droplet due to combustion.

Since, these studies involve evaporation (or burning), hence both liquid volume and surface area decrease with time. Therefore, the net effect of the local evaporation flux ($\dot{m''}$) and the total surface area ($A$) on the total evaporation rate ($\dot{m} = \dot{m''} \times A$) can be misleading. In order to dissociate the effect of total surface area on $\dot{m}$, these studies are performed where the liquid volume is kept constant artificially while still solving the remaining equations for the fluid dynamical system. 

\begin{figure}[h!]
   \centering
   \includegraphics[width=1\linewidth]{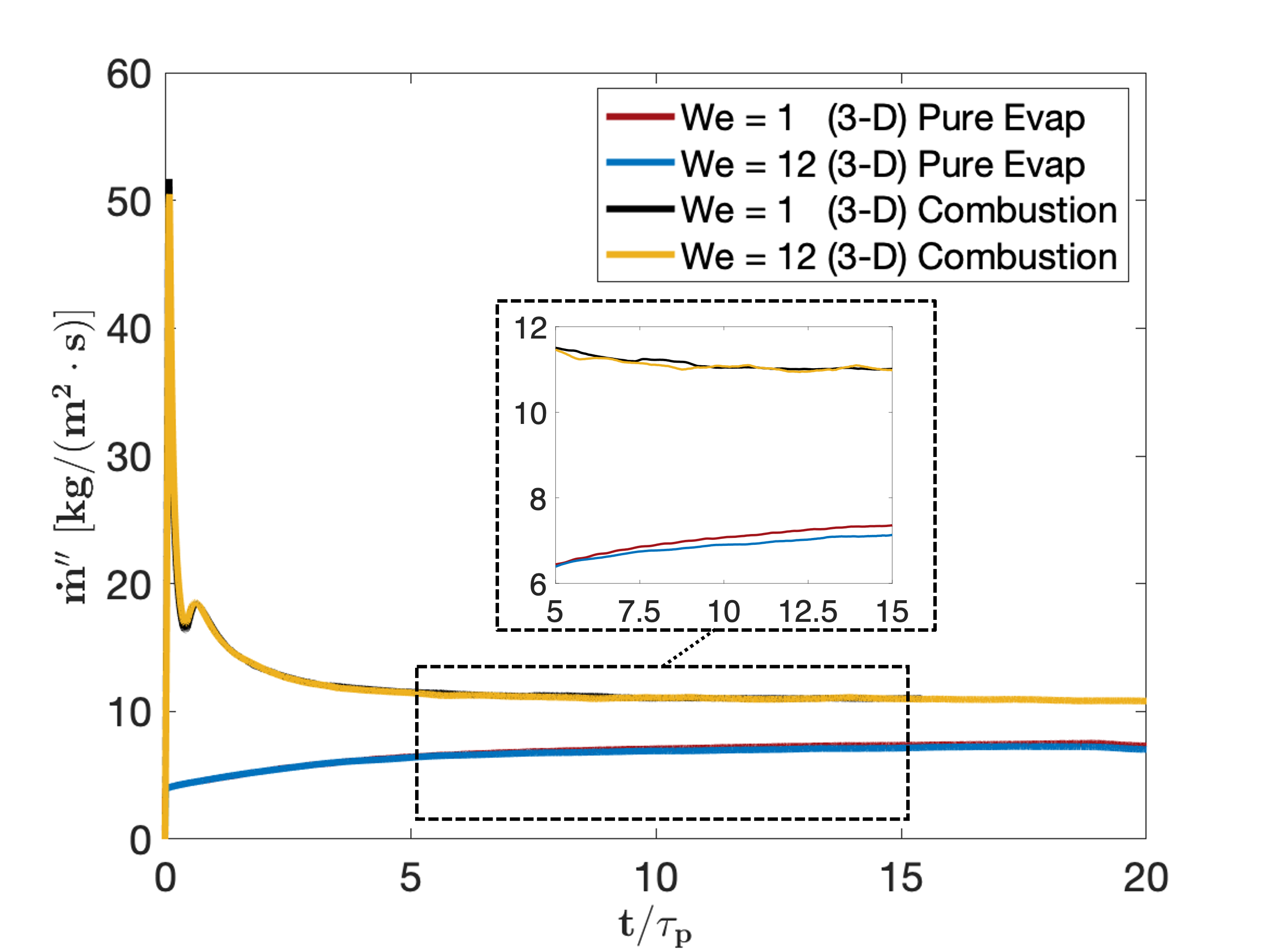} 
   \caption{Local evaporation rate $[kg/m^2 \cdot s]$ with $t/\tau_p$ at $Re_0=25$ }
   \label{fig:3d_weber_local_mdot}
\end{figure}

\begin{figure}[h!]
   \centering
   \includegraphics[width=1\linewidth]{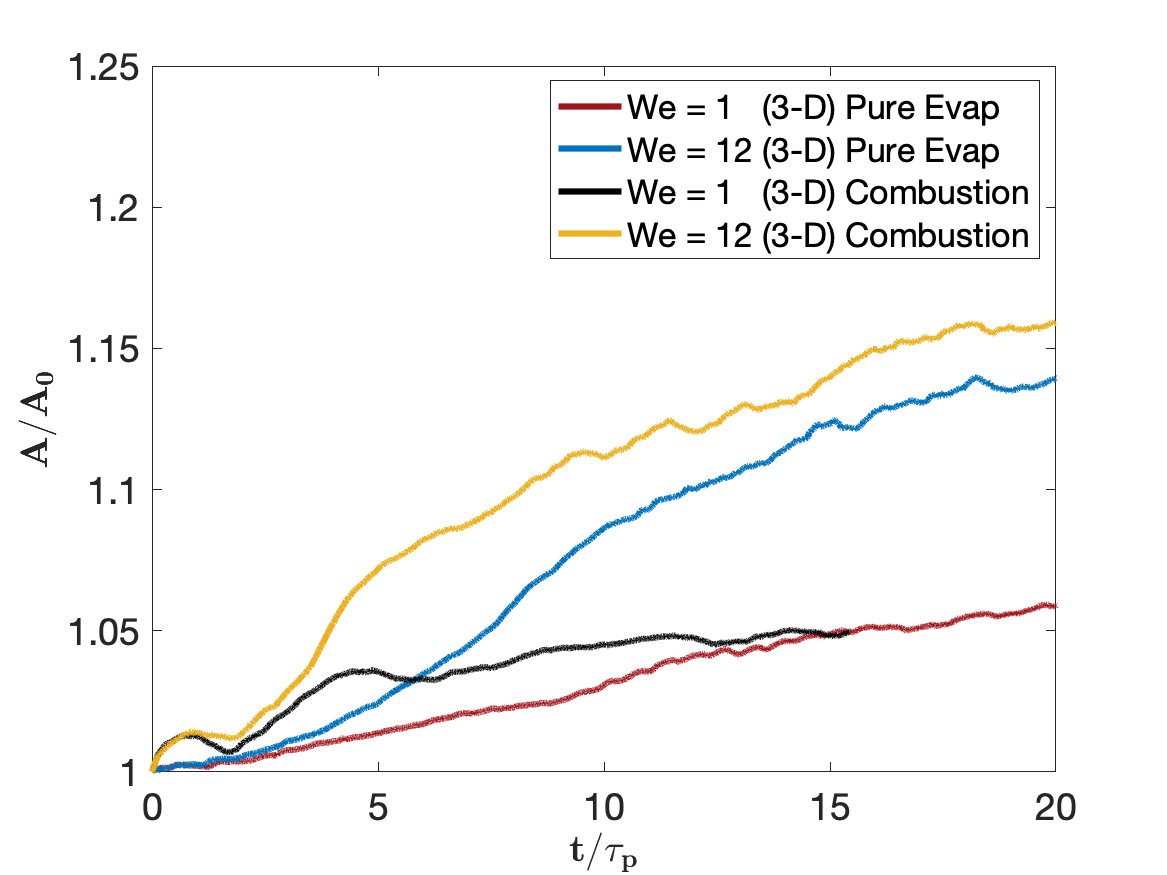} 
   \caption{Normalized total surface area with $t/\tau_p$ at $Re_0=25$ }
   \label{fig:3d_weber_area}
\end{figure}

To breakdown the contribution of evaporation flux and surface area in total evaporation rate, \cref{fig:3d_weber_local_mdot} and \cref{fig:3d_weber_area} show the time history of surface averaged local $\dot{m''}$ and normalized area $A/A_0$ respectively. In case of pure evaporation, the averaged value of $\dot{m''}$ for $We_0=1$ is nearly the same as $We_0=12$. The same is observed for combustion cases but with the higher magnitude of $\dot{m''}$. This suggest weak dependency of Weber number on averaged $\dot{m''}$ at this $Re_0$. 

Due to the presence of convective flow, the droplet deforms. This deformation can be quantified in terms of $A/A_0$ as shown in \cref{fig:3d_weber_area}. Here $A_0$ is initial droplet surface area. The results show a nearly linear increasing trend with time. This rise in $A/A_0$ is seen to be faster in the beginning of the evaporation which suggest higher deformations till $t/\tau_p \sim 10$. 
The slope of $A/A_0$ reduces at later time and $A/A_0$ reaches nearly a steady value. Moreover, $A/A_0$ is higher for $We_0=12$ when compared with $We_0=1$ for each case. For example: at a time instant $t/\tau_p \sim 7$, $We_0=12$ with combustion has $5.4\%$ higher total surface area when compared to $We_0=1$. This increase in area for $We_0=12$ pure evaporation case is $2.7\%$. One important observation in normalized area plot is that the trend of $A/A_0$ for a particular Weber number is same for both pure evaporation and combustion. They both differ by the magnitude of area ratio. 

\begin{figure} [h!]
\centering
\subfigure[$We_0=1$ : Pure Evaporation]{%
\label{fig:3d_evap_we_1}%
\includegraphics[width=0.45\linewidth]{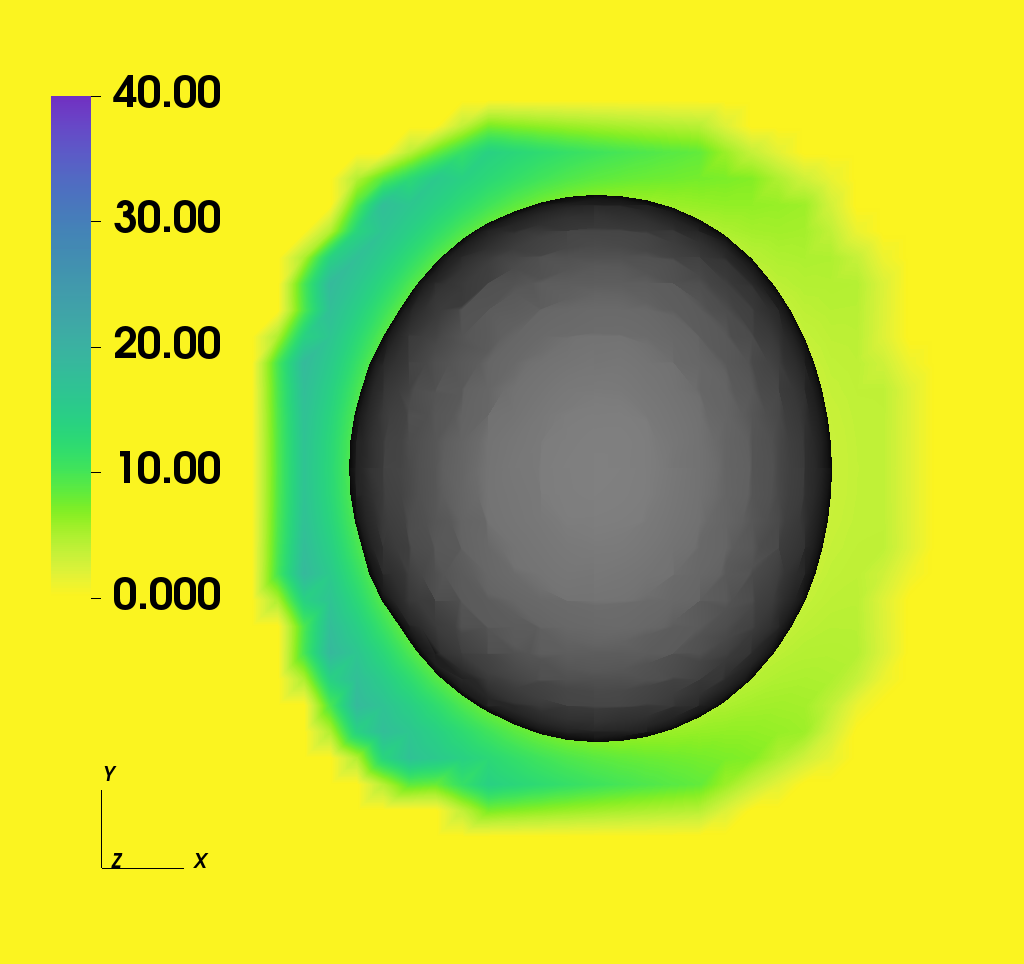}}%
\hspace{0.3cm}
\subfigure[$We_0=12$ : Pure Evaporation]{%
\label{fig:3d_evap_we_12}%
\includegraphics[width=0.45\linewidth]{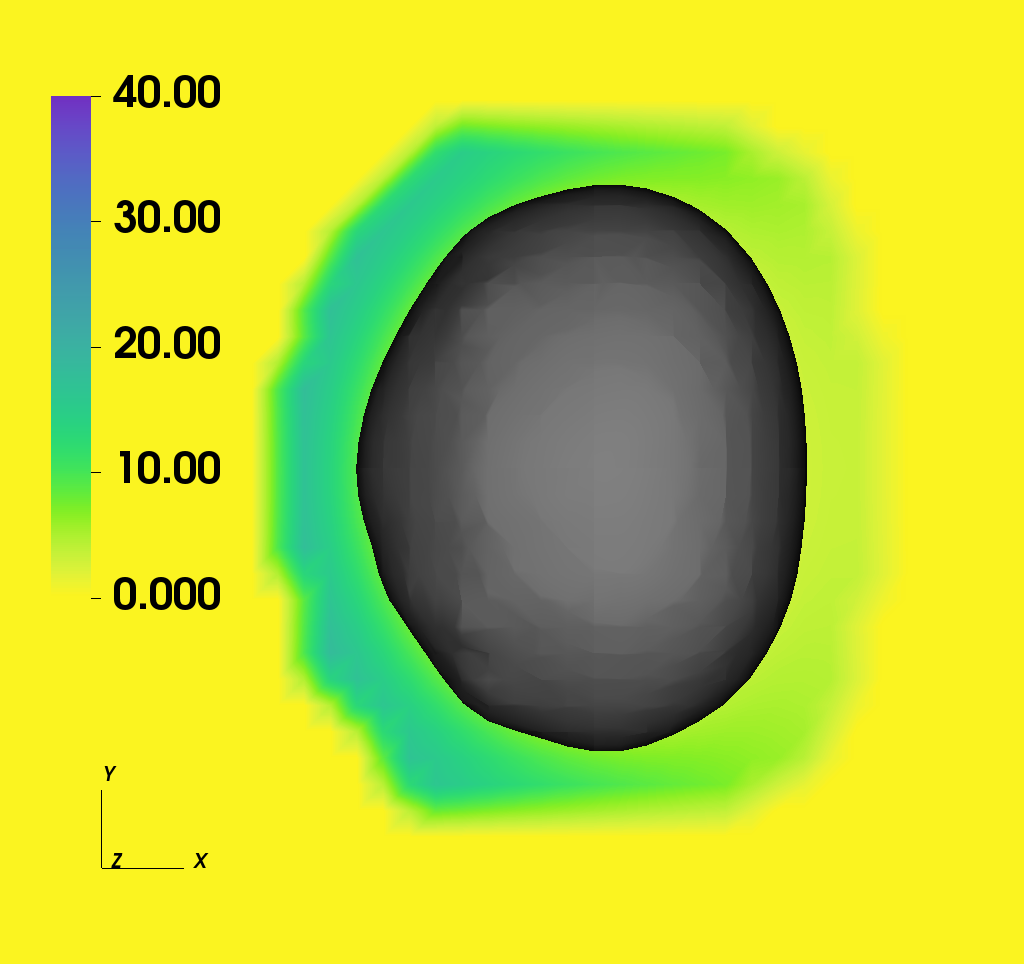}}%
\hspace{0.3cm}
\subfigure[$We_0=1$ : Combustion]{%
\label{fig:3d_cmbst_we_1}%
\includegraphics[width=0.45\linewidth]{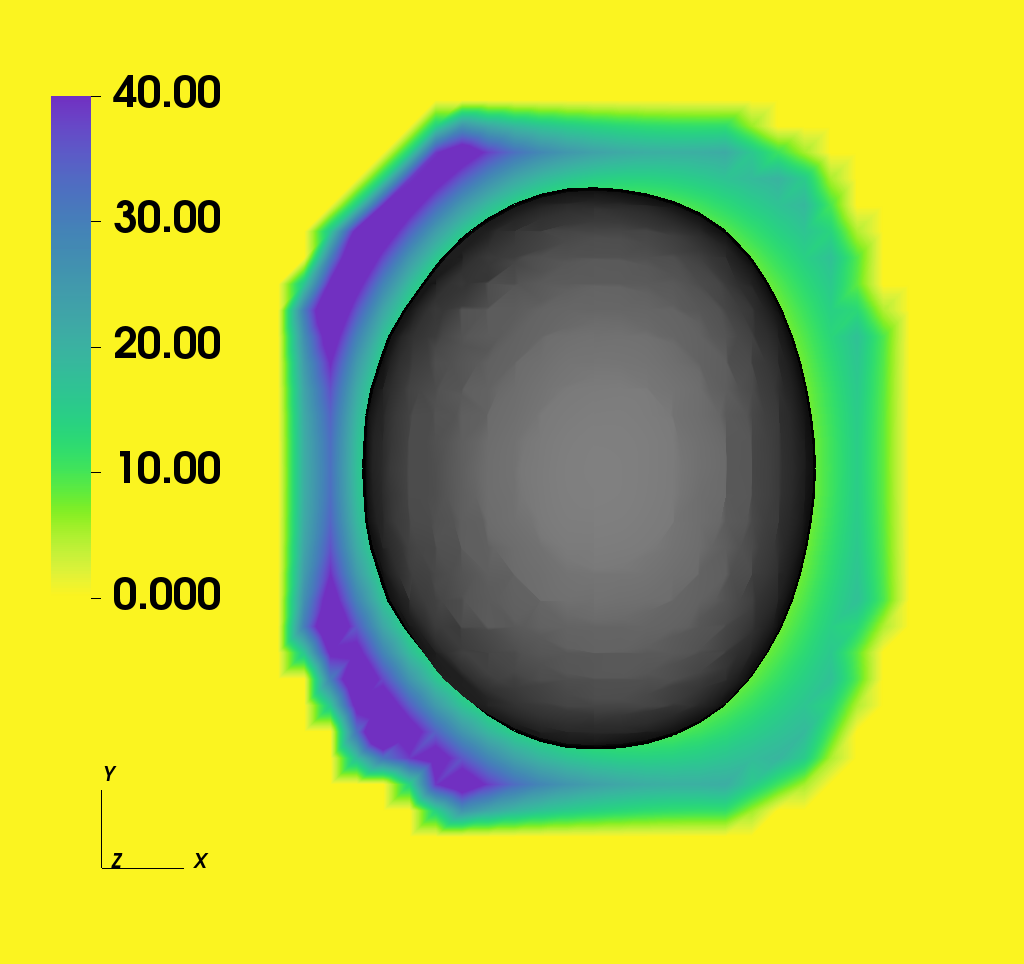}}%
\hspace{0.3cm}
\subfigure[$We_0=12$ : Combustion]{%
\label{fig:3d_cmbst_we_12}%
\includegraphics[width=0.45\linewidth]{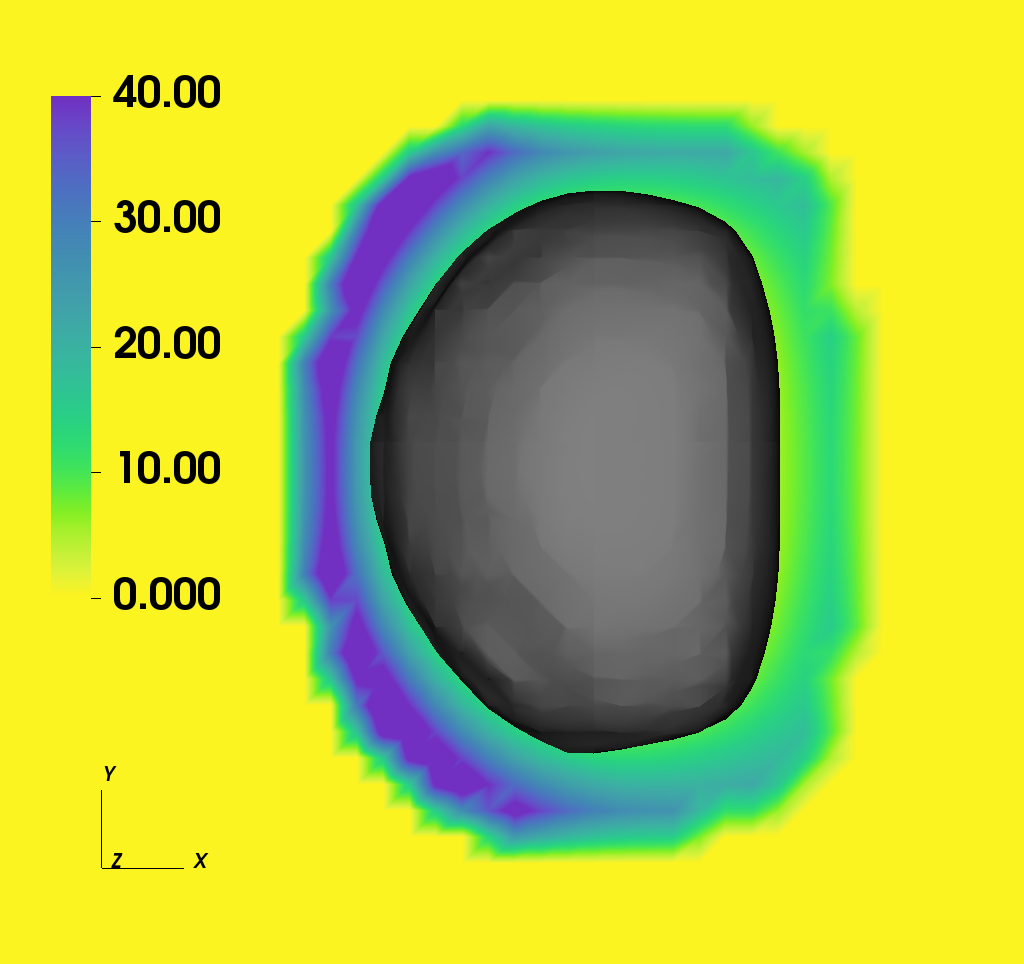}}%
\caption{Psueodcolors of local evaporation flux $[kg/(m^2 \cdot s)]$ at $t/\tau_p \sim 7$ at $Re_0=25$
}
\label{fig:3d_mdot}
\end{figure}

These trends in $A/A_0$ reflect into droplet shape change. Although surface averaged values of $\dot{m''}$ are nearly the same as seen in \cref{fig:3d_weber_local_mdot}, the droplet shape affects the distribution of local evaporation flux because of interaction between the droplet and inflow. To visualize these results, \cref{fig:3d_mdot} shows the pseudocolors  of local evaporation flux ($\dot{m''} \ [kg/(m^2 \cdot s)]$) at time instance $t/\tau_p \sim 7 $. The droplet shape at $We_0=1$ is close to an ellipsoid whereas it is more deformed at $We_0=12$ case for both pure evaporation and combustion. In addition to that, the droplet looks further deformed in $We_0=12$ combustion case (\cref{fig:3d_cmbst_we_12}) when compared to $We_0=12$ pure evaporation case (\cref{fig:3d_evap_we_12}). This suggests an interaction between droplet shape and combustion dynamics.

\begin{figure} [h!]
\centering
\subfigure[$We_0=1$ : Combustion]{%
\label{fig:3d_cmbst_we_1_curv}%
\includegraphics[width=0.45\linewidth]{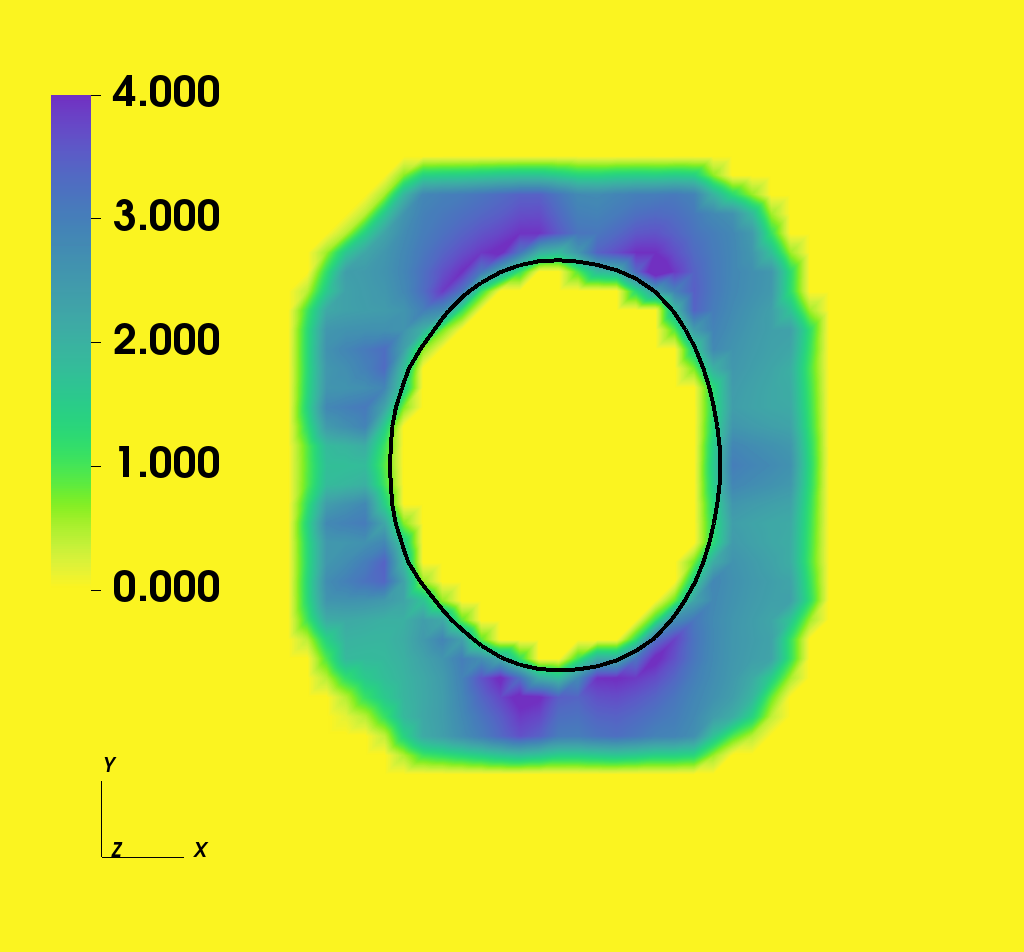}}%
\hspace{0.3cm}
\subfigure[$We_0=12$ : Combustion]{%
\label{fig:3d_cmbst_we_12_curv}%
\includegraphics[width=0.45\linewidth]{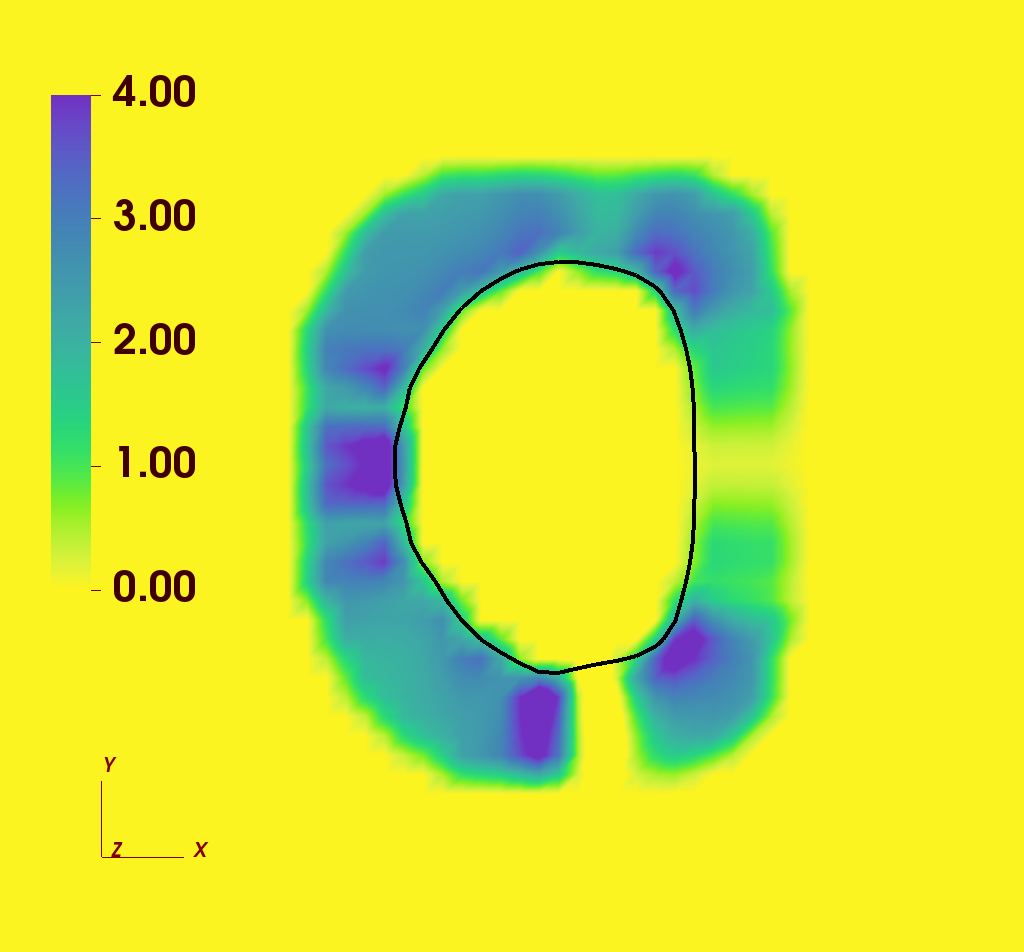}}%
\caption{Psueodcolors of normalized curvature at $t/\tau_p \sim 7$ at $Re_0=25$ \\ \textit{Black line marks the liquid-gas interface.}}
\label{fig:3d_curvature}
\end{figure}

Due to the convective flow over the droplet, the front of the droplet is exposed to hot incoming air. This leads to higher local flux in front of the droplet. Moreover, the distribution of $\dot{m''}$ in this region depends on the curvature of droplet. Now, while comparing the effect of Weber number for combustion cases in front of the droplet, higher $\dot{m''}$ is observed in $We_0=12$ due to the presence of higher curvature when compared to $We_0=1$ as shown in \cref{fig:3d_curvature}. The curvature is normalized using curvature of a sphere of diameter $d_0$. The values for $\dot{m''}$ measured near to the front stagnation point are $41.5 \ kg/(m^2 \cdot s)$ and $51.6 \ kg/(m^2 \cdot s)$ for $We_0=1$ and $We_0=12$ respectively. This observation is consistent with the previous studies showing that $\dot{m''}$ is directly proportional to the interface curvature at least in regions where the boundary layer is fully attached \cite{tonini2013},\cite{palmore_JHT_2022}. This was also observed in our previous work \cite{SETIYA2023104455}, \cite{Setiya2022Imece}.

\subsubsection{Effect of droplet shape on flame}
As the auto-ignition temperature for n-decane is $481 \ K$, for the given ambient conditions, the droplet starts burning and the envelope flame appears fairly early in time. This type of flame occurs due to faster reaction rates in comparison to convective transport. Initial Damk\"{o}hler number at $Re_0=25$ and for these ambient conditions is higher than critical value (${Da}_0 > 1.02$) as reported in \cite{wu2010}. One way of locating such flame is where stoichiometric combustion of the fuel and oxidizer occurs \cite{sirignano_1999}. This can be defined as a variable $\beta = Y_{\ce{F}} - Y_{\ce{O2}}/(A/F)_{st}$. Hence, $\beta=0$ marks the location of the flame. Another way of marking the flame location is to plot the contour of maximum species mass fraction for products such as $\ce{CO2}$ and $\ce{H2O}$. 

\begin{figure} [h!]
\centering
\subfigure[$We_0=1$]{%
\label{fig:3d_cmbst_we_1_gT}%
\includegraphics[width=0.7\linewidth]{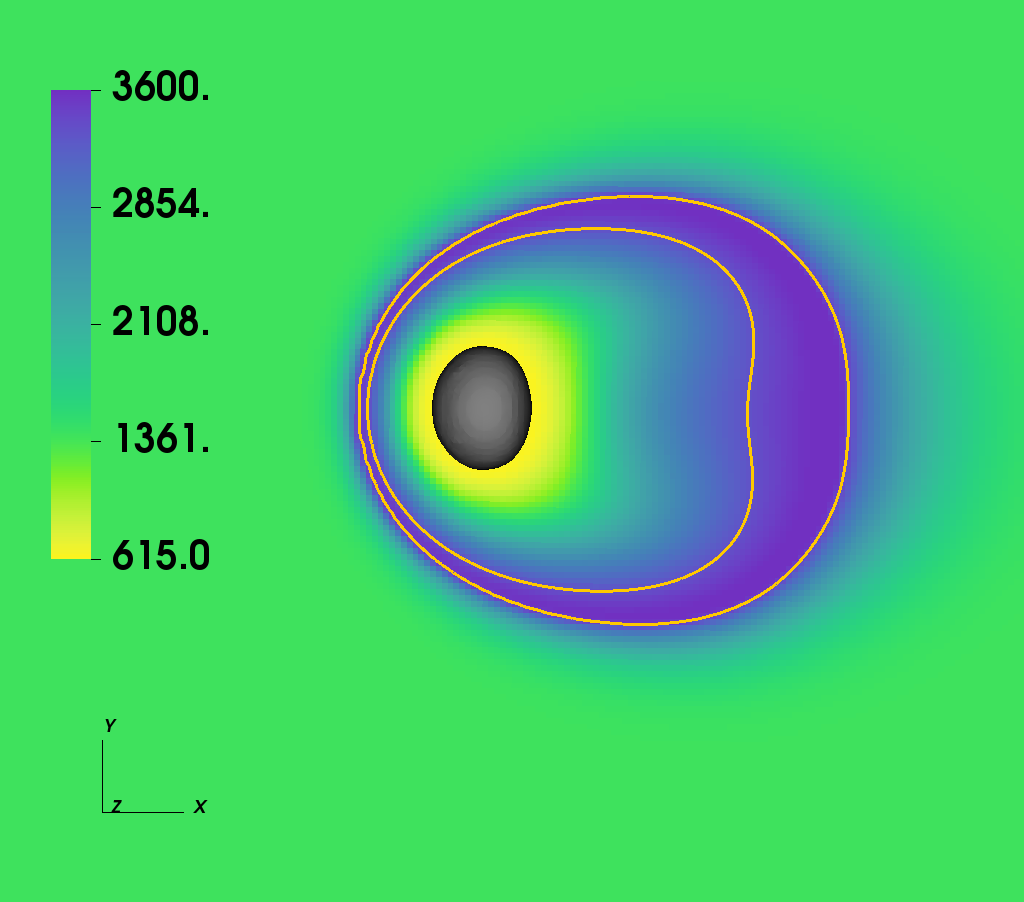}}%
\hspace{0.3cm}
\subfigure[$We_0=12$ ]{%
\label{fig:3d_cmbst_we_12_gT}%
\includegraphics[width=0.7\linewidth]{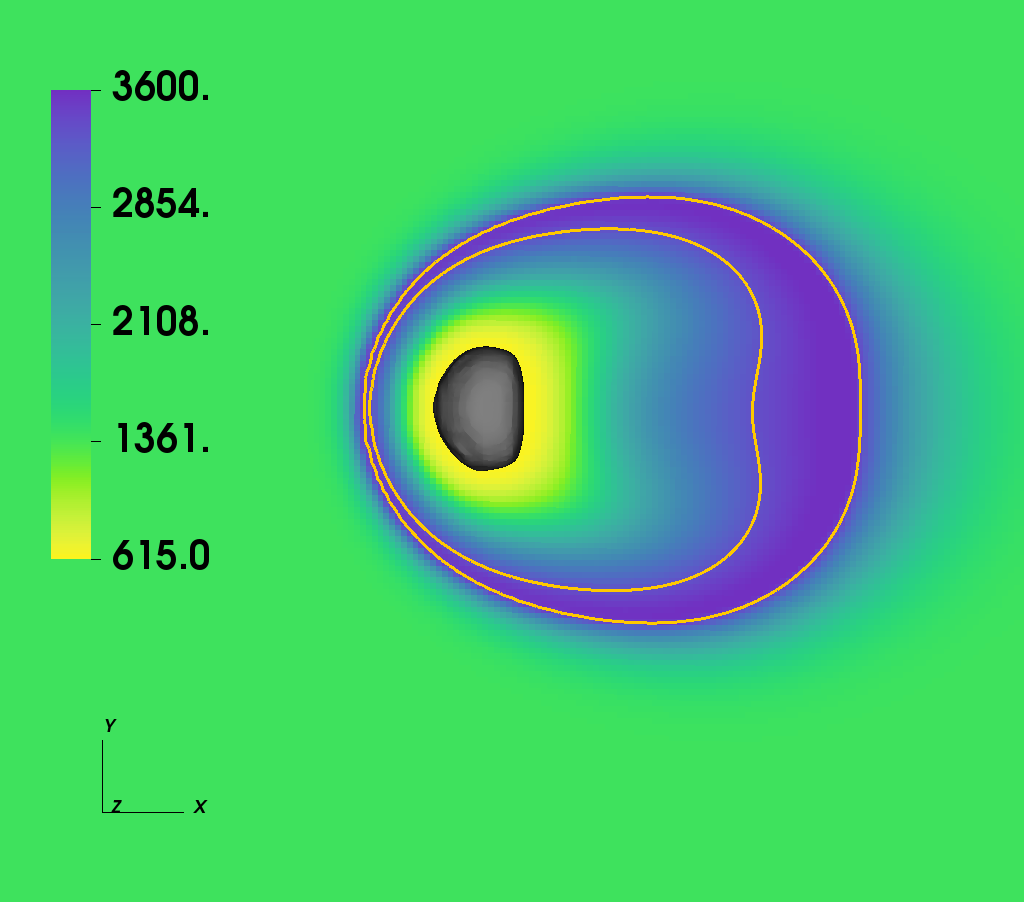}}%
\caption{Psuedocolors of gas temperature [$K$] at $Re_0=25$ at $t/\tau_p \sim 7$
\\
\textit{Yellow line marks the maximum $Y_{Co_2}$}}
\label{fig:3d_weber_gT}
\end{figure}

To evaluate the effect of Weber number on the combustion, \cref{fig:3d_weber_gT} shows the gas temperature around the droplet along with the contour of maximum species mass fraction of $\ce{CO2}$ ($max \ Y_{\ce{CO2}}$ - marked in yellow line). The flame is pushed towards the droplet and no longer stays as spherical diffusion flame. The gas temperature in the droplet vicinity is significantly lower in comparison to temperature in the flame region due to cooler fuel vapor. The region of maximum temperature lies where fuel vapor meets with oxidizer. This region is seen to coincide with maximum $Y_{\ce{CO2}}$ marked with yellow lines for each case. 

\begin{figure}[h!]
   \centering
   \includegraphics[width=0.7\linewidth]{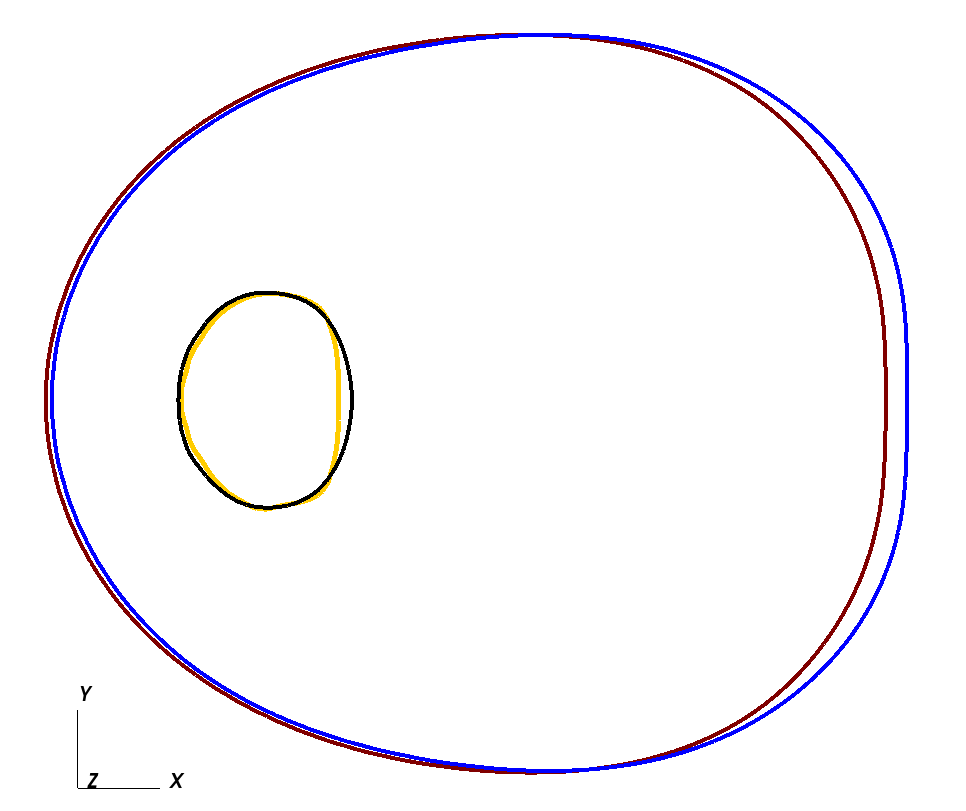} 
    \caption{ Sectional view of droplet and flame shape: $We_0=1$ (black) vs $We_0=12$ (orange), $Re_0=25$ \\
    \textit{Maroon line marks flame for $We_0=1$ and blue line marks flame for $We_0=12$}}
    \label{fig:3d_weber_shape}
\end{figure}

In terms of comparison between different Weber numbers, same magnitude of maximum gas temperatures as well as maximum $Y_{\ce{CO2}}$ are observed despite the changes in $\dot{m''}$ due to shape change in the front. In addition to this, the flame type is still the envelope non-spherical flame for given ambient conditions. Moreover, the contour for maximum $Y_{\ce{CO2}}$ and distribution of gas temperature are seen to be qualitatively same for both Weber numbers.
To compare the flames further  \cref{fig:3d_weber_shape} show an overlay of  z-section of droplet shape and the flame marked by $\beta=0$ for different Weber numbers. 

The flame base distance (measured from the droplet center to $\beta=0$ along negative x-direction) and flame stand-off distance (measured from droplet center to $\beta=0$ along positive y-direction) are observed to be same for both Weber numbers. However, the flame length behind the droplet (measured from the center of the droplet to end of the flame in flow direction) is seen to be longer for $We_0=12$. The evaporation in rear region of the droplet is dependent on the flow interaction in wake region which impacts the flame length behind the droplet. 
The flow velocity measured at the end of flame is found to be higher in $We_0=12$. This leads to the flame location to  be pushed further downstream. The values of velocity and non-dimensional flame length are tabulated in \cref{table:2}. 

\begin{figure} [h!]
\centering
\includegraphics[width=0.7\linewidth]{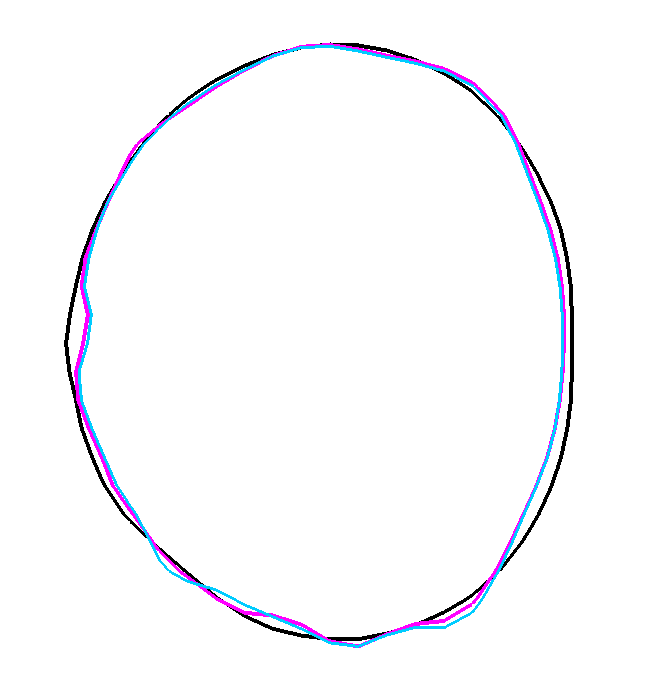}
\caption{Sectional view of droplet shapes at different Weber number at $t/\tau_p \sim 10$, \\
$We_0=1$ (Black), $We_0=16$ (Magenta), $We_0=20$ (Cyan) }
\label{fig:all_weber_cmbst}
\end{figure}

\begin{table}[h!]
\caption{Measurement of flow velocity $v$ and normalized flame length $L^*$ at $Re_0=25$ }
\label{table:2}
\begin{center}
\begin{tabular}{ |p{3cm}|p{2cm}|p{2cm}| } 
\hline
Case & $v \ [m/s]$ & $L^* = L/d_0$\\ 
\hline
$We_0=1$  & 1.65 & 3.06\\ 
$We_0=12$ & 1.78 & 3.17\\ 
\hline
\end{tabular}
\end{center}
\end{table}
\textcolor{black}{Overall, combustion at $Re_0=25$ shows little sensitivity to the Weber number. Further analysis, however, demonstrates that in this case none of the droplets deform significantly. To further explore this effect, we introduce simulations at $We_0=16\text{ and }20$ The cross-section of the droplets at different Weber numbers are overlaid in \cref{fig:all_weber_cmbst} to compare the droplet shape. These results show that the droplet dimension perpendicular to flow is nearly the same for all Weber number cases. Moreover, droplet break-up is not observed even at $We_0=20$ when the droplet is burning. This seems to contradict previous literature on the droplet Weber number effect, which predicts increasing deformation with increasing Weber number. In particular, Weber number $We_0=12$ was reported as a critical $We_0$ for the onset of breakup in previous literature such as \cite{loth2006,KEKESI20141,hinze1955}.}

\begin{figure} [h!]
\centering
\includegraphics[width=1\linewidth]{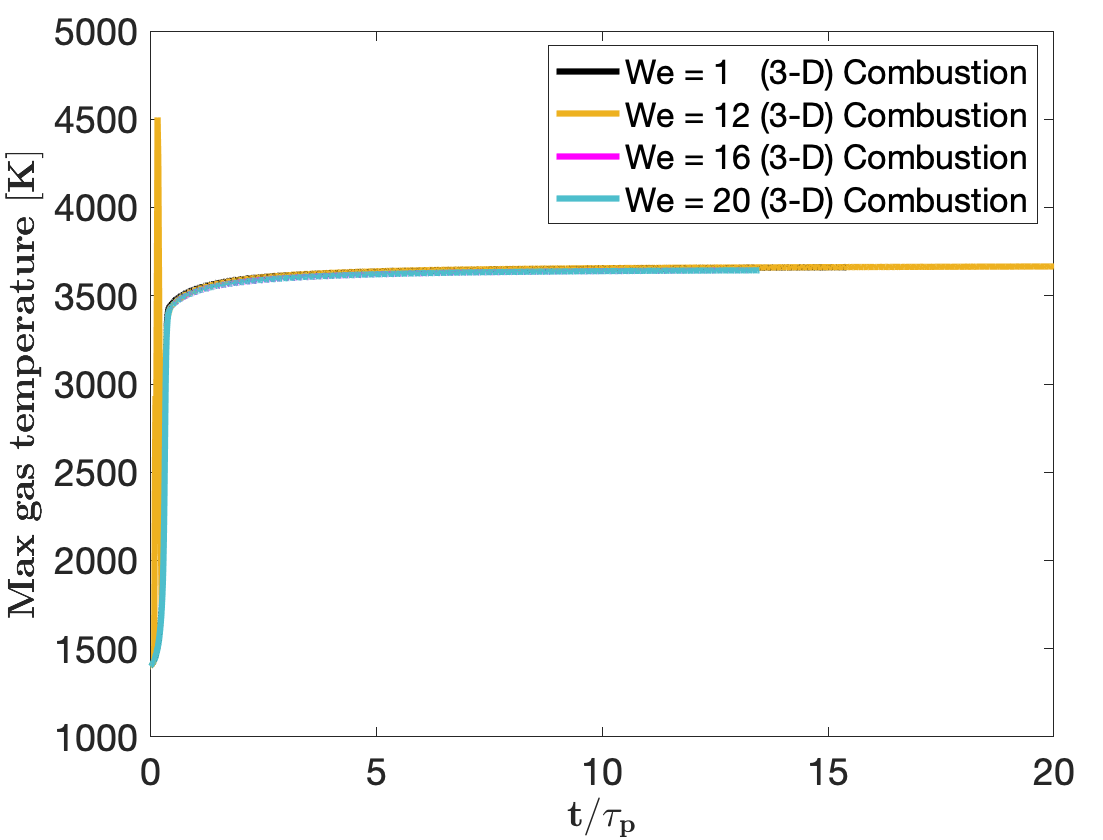}
\caption{Comparison of gas temperature for different Weber numbers at $Re_0=25$}
\label{fig:3d_we_comp}
\end{figure}

\textcolor{black}{As a result of this lack of deformation, the results all show surprisingly similar combustion behavior as $We_0=1$ in terms of gas temperature and species mass fraction of $CO_2$. The time history of gas temperature for the new cases is shown in \cref{fig:3d_we_comp} which suggests nearly the same gas temperature for all cases. The spatially averaged gas temperatures and species mass fraction of $CO_2$ (not shown here) are seen to be closely the same for all Weber numbers.} 

\begin{figure} [h!]
\centering
\label{fig:shape_2}%
\includegraphics[width=0.7\linewidth]{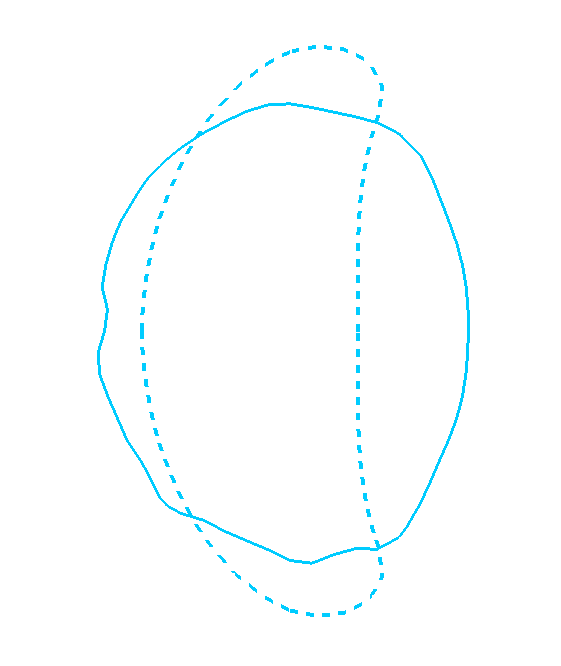}
\caption{Sectional view of droplet shape at $We_0=20$ at $t/\tau_p \sim 10 $, Cyan: with combustion, Cyan dashed: without combustion  }
\label{fig:shape_comp}
\end{figure}

\textcolor{black}{To investigate this phenomenon further, we simulated non-combusting, non-evaporating droplets at $We_0=16,\ 20$ and compared them against the corresponding combusting cases. A cross-section of the droplet for $We_0=20$ cases is compared in \cref{fig:shape_comp} at time instant $t/\tau_p \sim 10$. Smaller deformations are seen for burning droplet at the same Weber number. Moreover, the droplet break up is observed at time $t/\tau_p \sim 87, \ 81$ for $We_0= 16$ and $We_0=20$ respectively for non-combusting droplet. This clearly points toward a complicated interaction between combustion and droplet shape as the droplet never reaches to its maximum deformation while burning. It also suggests that the droplet deformation is inhibited due to higher evaporation rates in combustion case. This can also be understood from the jump condition in \cref{eq:PJ}, where the surface tension must balance the pressure difference across the interface and the evaporation flux. For non-evaporating cases, the balance is purely between surface tension and pressure. A large curvature ($\kappa$) is required for the surface tension to balance the aerodynamic pressures from the flow around the droplet. Hence, larger deformation results to create this curvature. However, when evaporation is introduced, it works in conjunction with the surface tension to balance the pressure jump. Therefore such large deformations are not required. This is related in concept to established literature that shows the droplet break-up does not only depend on Weber number. Other conditions such as density ratio, viscosity ratio, Ohnesorge number, and ambient gas velocity affect the deformation, break-up time, and break-up mode \cite{PILCH1987741, Villermaux2009}. This current paper suggests that the evaporation rate is also an important factor in predicting droplet deformation and break-up.}

\subsubsection{Internal Circulation}
\textcolor{black}{It may be expected that the rate of internal circulation inside the droplet can affect the evaporation and combustion properties. In this paper, the droplet is pre-heated, so internal circulation only affects the evaporation and combustion via changing the boundary layer development in and around the droplets. This effect was demonstrated in \cite{palmore_scitech2022} which investigated the evaporation of multicomponent jet fuel surrogate.}

\begin{figure} [h!]
\centering
\subfigure[$We_0=1$ : Combustion]{%
\label{fig:3d_cmbst_we_1_liq_vel_vec}%
\includegraphics[width=0.7\linewidth]{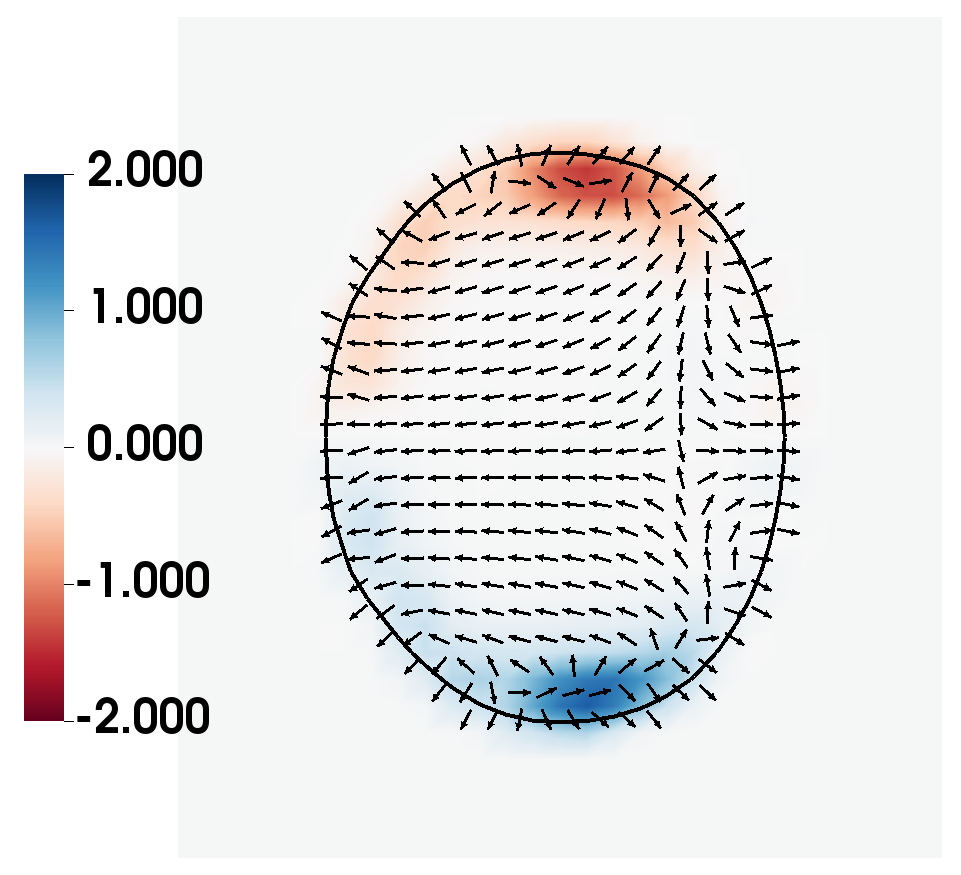}}%
\hspace{0.3cm}
\subfigure[$We_0=12$ : Combustion]{%
\label{fig:3d_cmbst_we_12_liq_vel_vec}%
\includegraphics[width=0.7\linewidth]{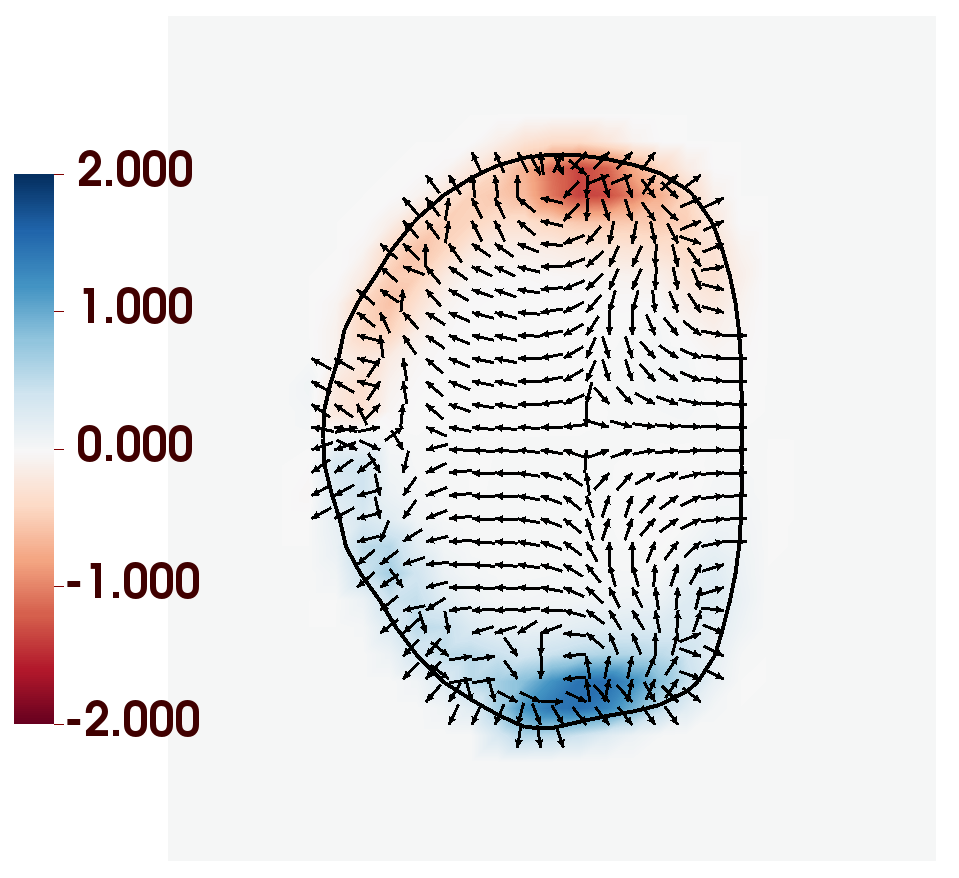}}%
\caption{Normalized vorticity with vectors of normalized liquid velocity at $t/\tau_p \sim 7$ at $Re_0=25$ \\ \textit{Black line marks the liquid-gas interface.}}
\label{fig:3d_liq_vel_vec}
\end{figure}

\textcolor{black}{Past studies on the effect of internal circulation \cite{LAW1977605,yushu_scitech,yushu_PRF2022} showed that the internal circulation can be predicted by the density ratio of two fluids ($\rho_l/\rho_g$). As the density ratio increases, the magnitude of internal circulation is observed to be decreasing. Therefore, higher density ratio cases demonstrated weaker internal circulation. The density ratio for the presented work is $\rho_l/\rho_g \approx 60$, so internal circulation is expected to be negligible. To demonstrate the internal motion for this study, \cref{fig:3d_liq_vel_vec} shows vectors of normalized liquid velocity overlaid on normalized vorticity inside droplet. The initial inflow velocity $U_0$ and initial droplet diameter $d_0$ are used for the normalization. A counter-rotating vortex pair is observed near the top and bottom of droplet in both cases. However, the overall magnitude of normalized liquid velocity remains low ($\le 7.5 \% U_\infty$) in both cases. Therefore, the effect of such internal motion is not expected to be significant. This agrees well with the observations in \cite{yushu_PRF2022}.
}


\subsection{Comparison of 3-D vs 2-D Results}\label{sec:2d-3d}
A common strategy, particularly for simulations, is to develop and test models for multiphase flows such as boiling, evaporation in 2-D \cite{Shao2018,GIBOU2007536,TANGUY20141}. This is due to the significant reduction in computational cost associated to this change in dimensionality. However, this leads to the question of how useful such validations are. Can 2-D studies be used to gain useful trend information about 3-D flows, or are the physics of three dimensionality inherently important to the study of combustion?
This can be done by interrogating if 2-D Cartesian analysis sufficient for simulating the droplet evaporation and combustion. Hence, as an additional insight to this research question, this section discusses the qualitative and quantitative differences between 2-D and 3-D results for combustion at $We_0=12$. The reason for selecting this Weber number is to analyze the differences for highly deformed shape, making it a more conservative test than the lower Weber number case.

\begin{figure}[h!]
   \centering
   \includegraphics[width=1\linewidth]{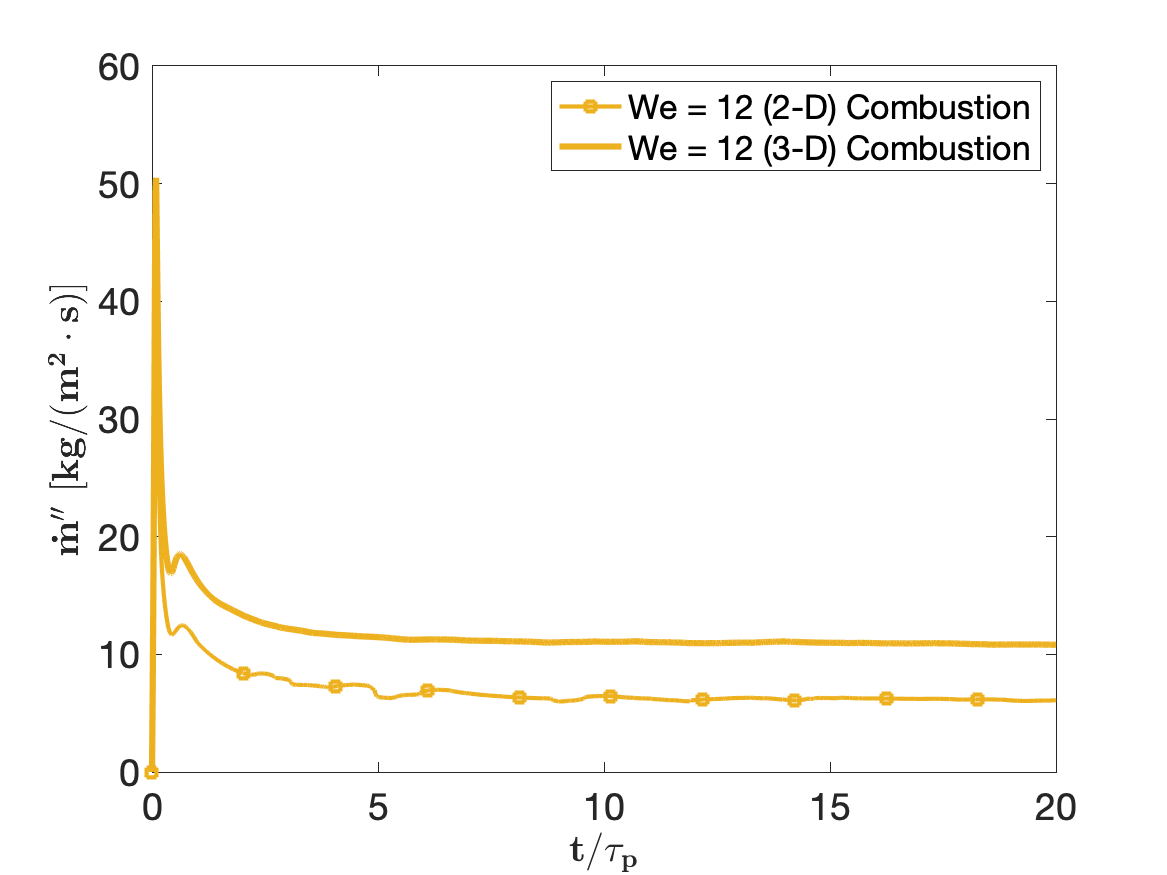} 
   \caption{2-D vs 3-D : Local evaporation flux $ [kg/(m^2 \cdot s)] $ with $t/\tau_p$ at $We_0=12$ , $Re=25$ }
   \label{fig:2d_3d_we_12_local_mdot}
\end{figure}

To evaluate 2-D vs 3-D, the comparison of local evaporation flux ($\dot{m''}$), the area evolution ($A/A_0$), the flame stand-off distance and flame shape is presented in this section.
\Cref{fig:2d_3d_we_12_local_mdot} shows the comparison between 2-D and 3-D for combustion at $We_0=12$ for the given conditions. The surface-averaged $\dot{m''}$ is observed to be higher in 3-D when compared to 2-D. At $t/\tau_p \sim 15$ is $ 42.5 \%$ lower in 2-D (thin yellow line with markers) when compared to 3-D (thick yellow line).

\begin{figure} [h!]
\centering
\subfigure[$We_0=12$ : 2-D Combustion]{%
\label{fig:2d_evap_gT}%
\includegraphics[width=0.7\linewidth]{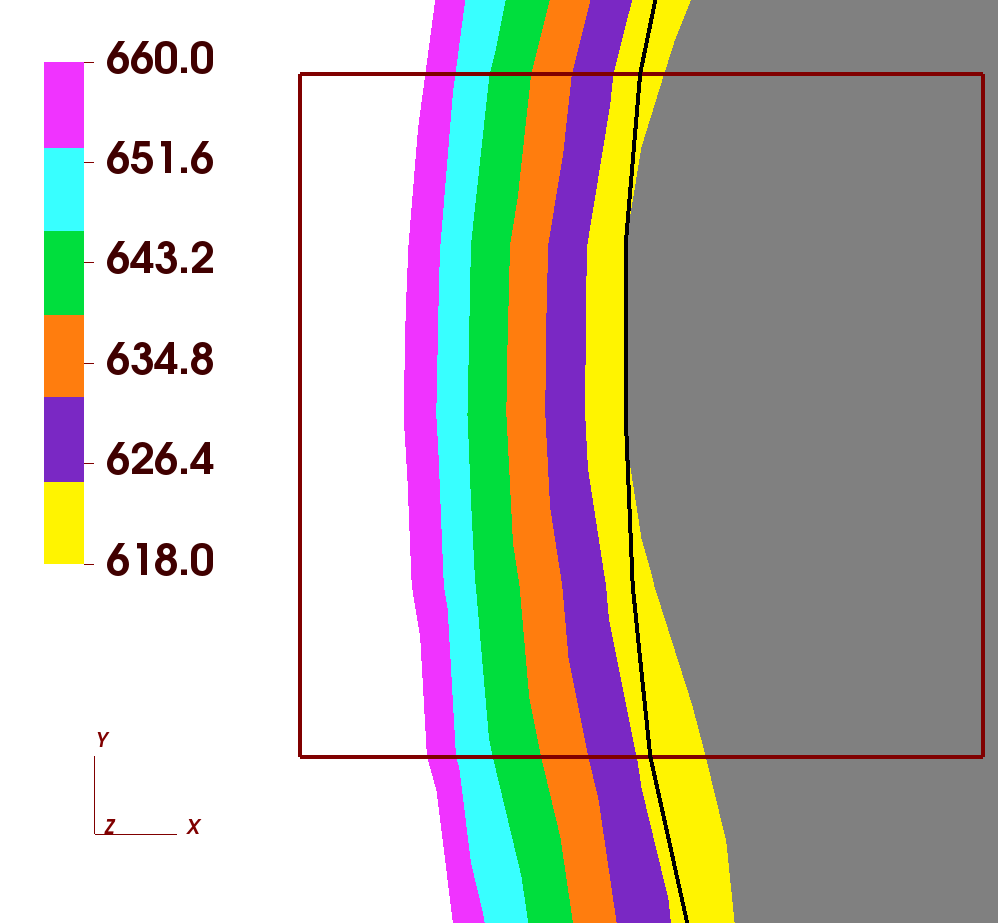}}%
\hspace{0.3cm}
\subfigure[$We_0=12$ : 3-D Combustion]{%
\label{fig:3d_evap_gT}%
\includegraphics[width=0.7\linewidth]{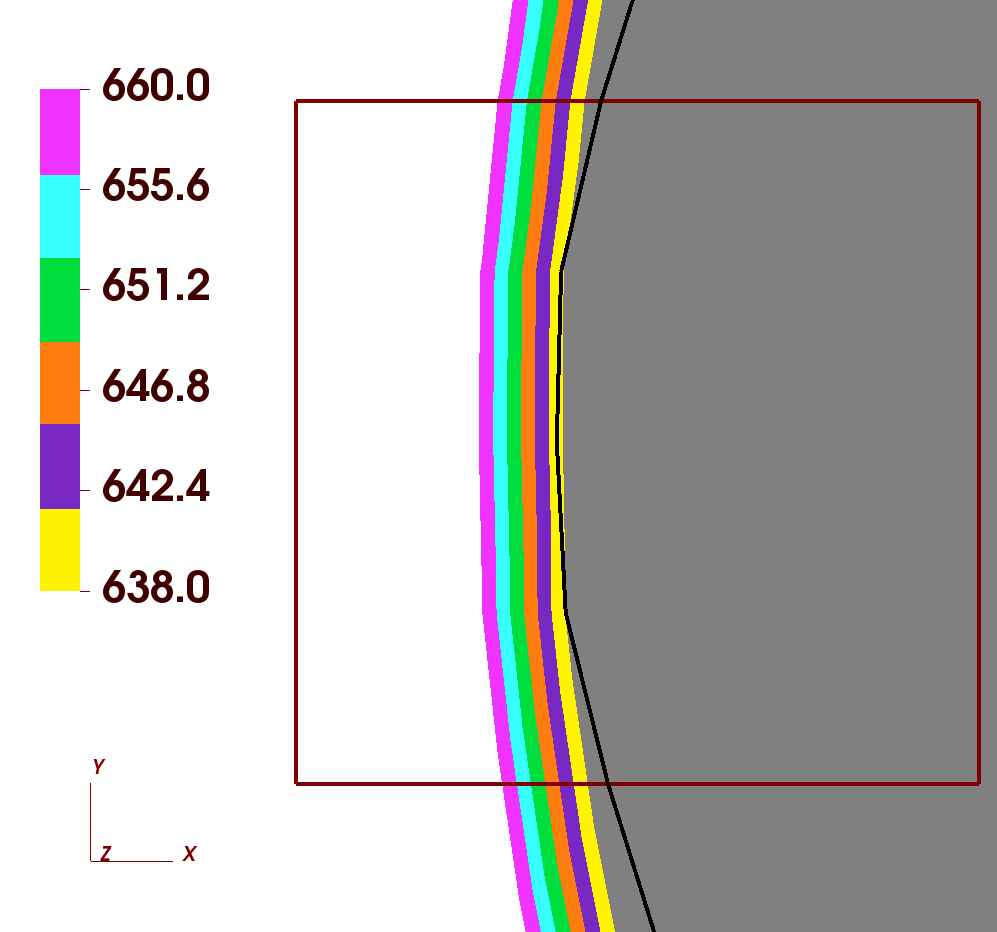}}%
\caption{ 2-D vs 3-D : Gas Temperature $[K]$ near L-G interface for combustion at $t/\tau_p \sim 7$, $We_0=12$, $Re_0=25$ \\
The maroon box is of size $l \times w = 20 \ \mu m \times 20 \  \mu m$}
\label{fig:2d_3d_evap_gT}
\end{figure}

This higher mass flux in 3-D case is associated with the higher temperature gradients near to the liquid-gas interface. Such temperature gradients near the front stagnation point are shown in \cref{fig:2d_3d_evap_gT} 
where the thickness of this region of thermal gradient is lower in \cref{fig:3d_evap_gT} than \cref{fig:2d_evap_gT}, hence gradients are higher in 3-D. 

\begin{figure} [h!]
\centering
\subfigure[2-D : $We_0=12$ Combustion]{%
\label{fig:2_2d_cmbst_we_12_curv}%
\includegraphics[width=0.45\linewidth]{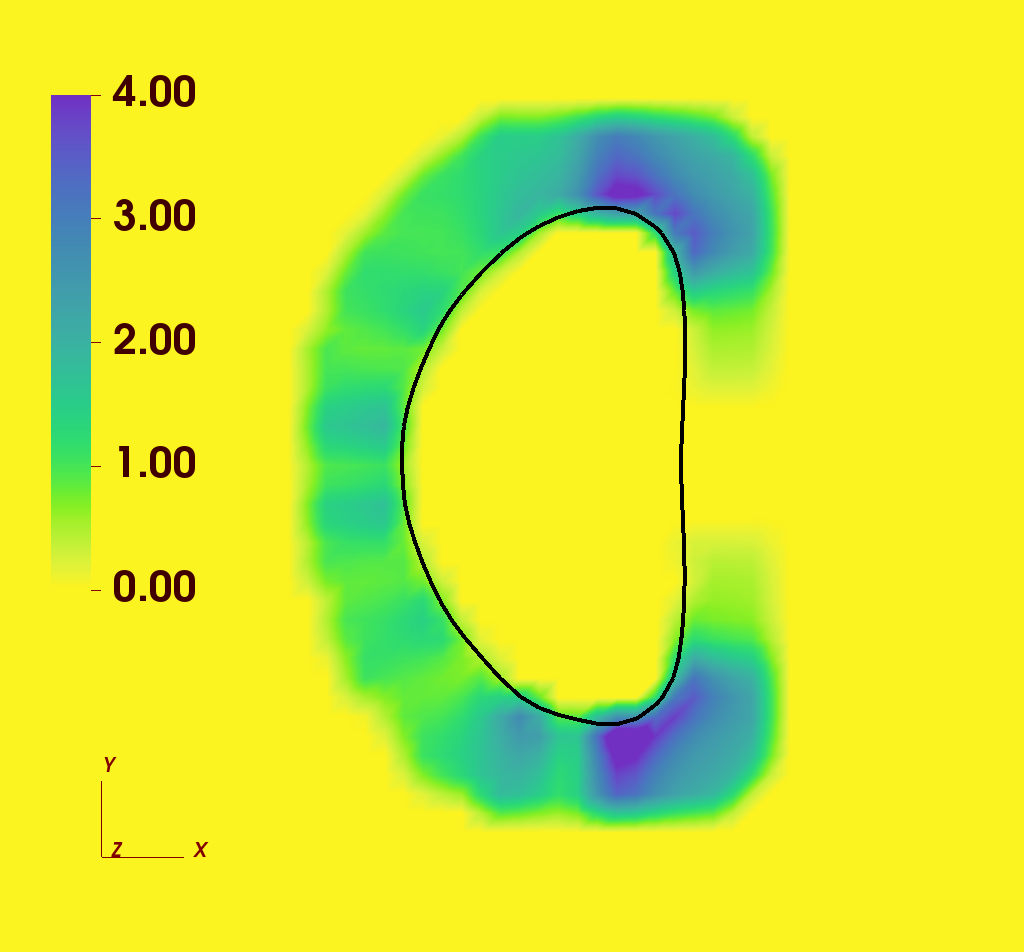}}%
\hspace{0.3cm}
\subfigure[3-D : $We_0=12$ Combustion]{%
\label{fig:1_3d_cmbst_we_12_curv}%
\includegraphics[width=0.45\linewidth]{3d_cmbst_we_12_curvature_t_0.00012s_v0.png}}%
\caption{2-D vs 3-D : Non-dimentional curvature at $t/\tau_p \sim 7 $, $We_0=12$ , $Re=25$ }
\label{fig:2d_3d_curvature}
\end{figure}

This region of higher temperature gradient in 3-D case is seen to coincide with the higher curvature near the windward stagnation point as shown in \cref{fig:1_3d_cmbst_we_12_curv} when compared to \cref{fig:2_2d_cmbst_we_12_curv}. The curvature is non-dimensionalized using the initial droplet diameter. The peak curvature in the stagnation region in 3-D is seen to be about twice that in 2-D, which agrees with trends for spherical/circular droplets in 3-D/2-D.  This higher curvature results in a higher local evaporation flux (\cref{fig:2d_3d_we_12_local_mdot}), consistent with the results from ~\cite{SETIYA2023104455}.

\begin{figure}[h!]
   \centering
   \includegraphics[width=1\linewidth]{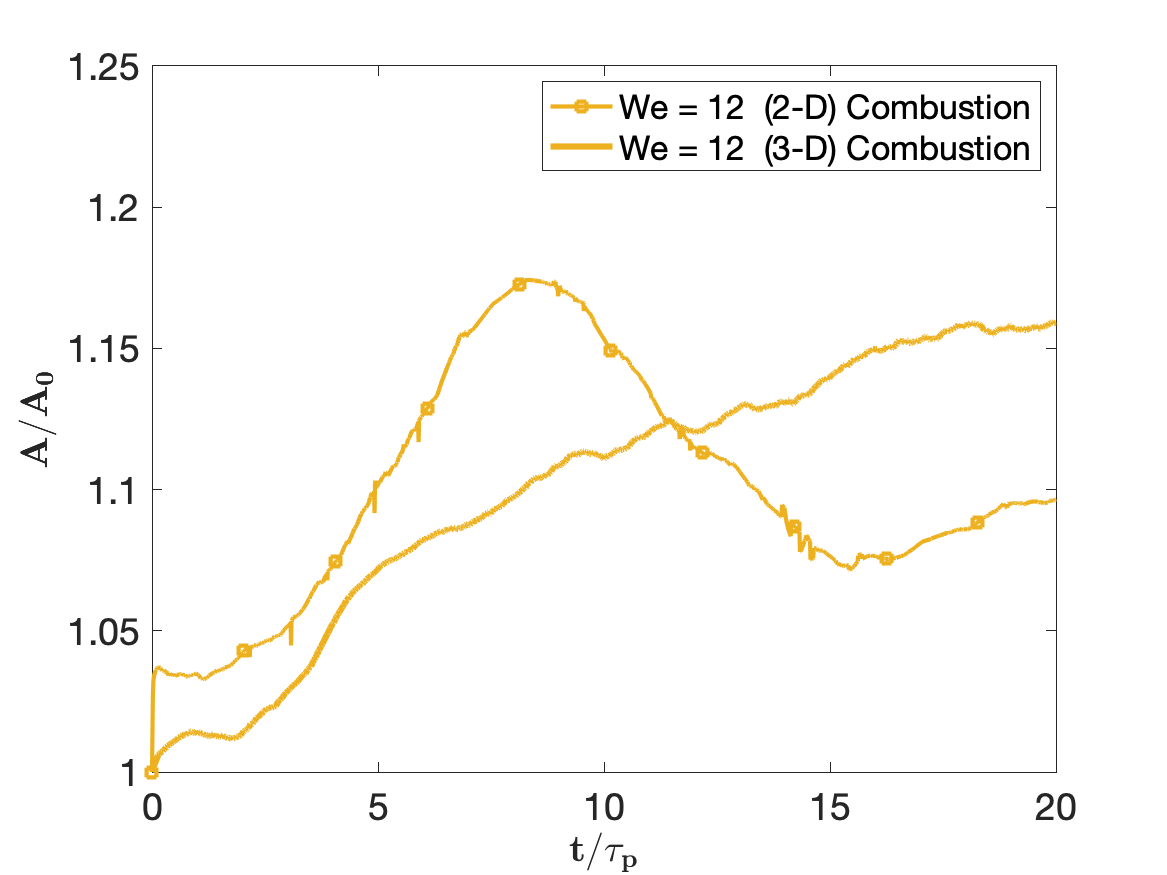} 
   \caption{2-D vs 3-D : Normalized total surface area with $t/\tau_p$ at $We_0=12$ , $Re=25$ }
   \label{fig:2D_3D_we_12_area}
\end{figure}

\begin{figure} [h!]
\centering
\subfigure[$We_0=12$ : 2-D Combustion]{%
\label{fig:2d_cmbst_vectors}%
\includegraphics[width=0.7\linewidth]{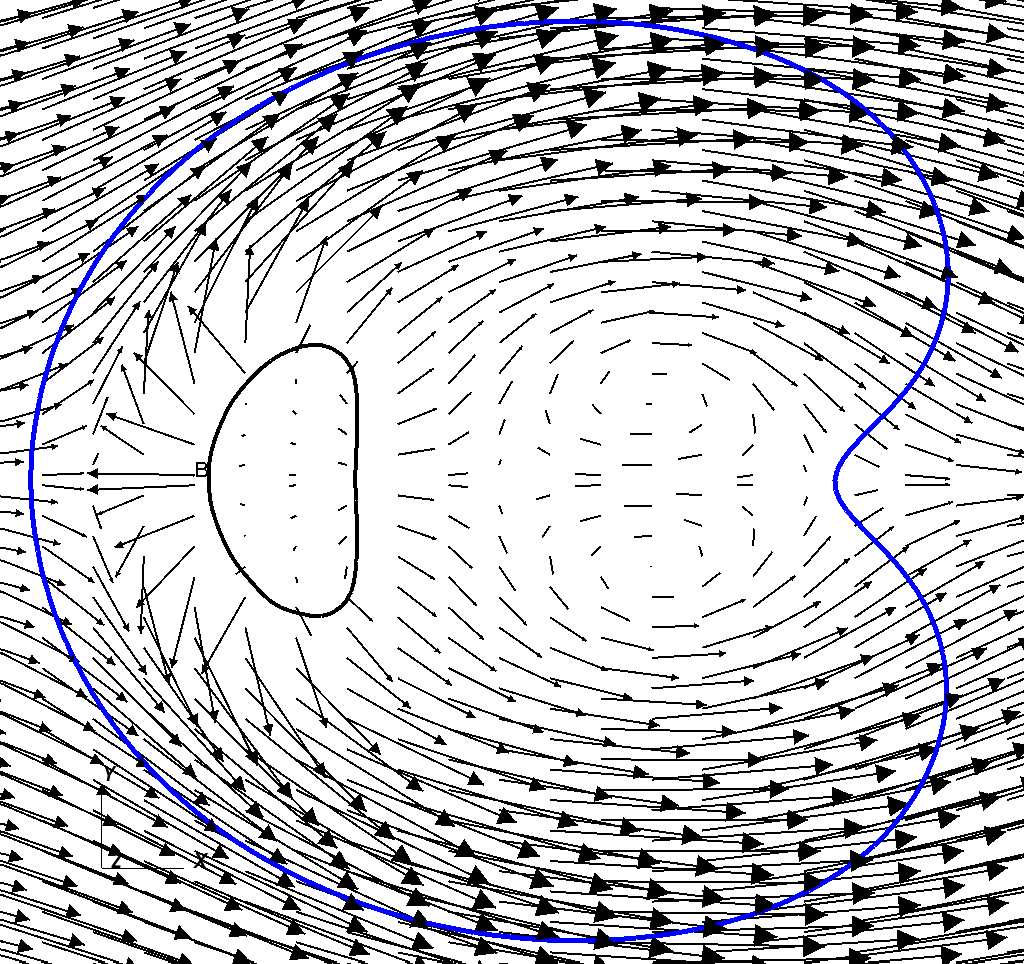}}%
\hspace{0.3cm}
\subfigure[$We_0=12$ : 3-D Combustion]{%
\label{fig:3d_cmbst_vectors}%
\includegraphics[width=0.7\linewidth]{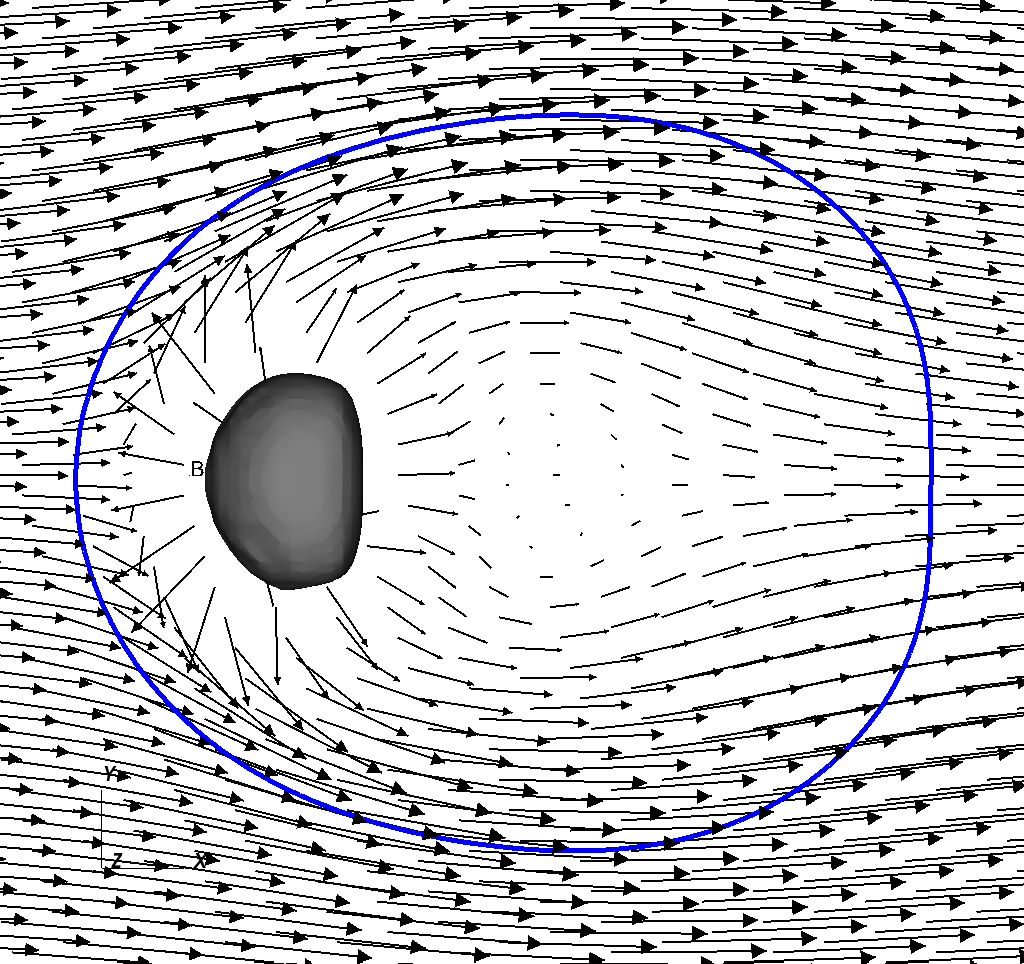}}%
\caption{ 2-D vs 3-D : Velocity vector field for combustion at $t/\tau_p \sim 7$, $Re_0=25$
\\ \textit{blue line marks the flame}}
\label{fig:2d_3d_cmbst_vectors}
\end{figure}

\Cref{fig:2D_3D_we_12_area} shows the comparison of $A/A_0$ for combustion at $We_0=12$, $Re_0=25$. $A/A_0$ in 3-D shows nearly linear increase in $A/A_0$ till $t/\tau_p \sim 12-15$ and tends to stabilize at a steady value at later time. However, 2-D results show the oscillations in $A/A_0$. These oscillations in $A/A_0$ continue till $t/\tau_p \sim 10-15$ and then reach nearly steady value. As observed from these results, the trends of normalized surface area indicates that the selection of droplet geometry has a clear impact on the way it deforms.
Furthermore, 2-D case has higher initial deformation when compared to 3-D case, however the long term trend is opposite. 
This comparison of shape is shown in \cref{fig:2d_3d_cmbst_vectors}. 


To evaluate the difference between 2-D and 3-D on combustion at $We_0=12$, \cref{fig:2d_3d_cmbst_vectors} shows the flame shape and flow field around the droplet. 
This demonstrates that although some aspects of the combustion are altered, the fundamental chemical behaviors between 2-D and 3-D droplets remains largely unchanged.
However, an intriguing observation is the flame shape at the downstream of the droplet. In 2-D, a stronger vortical region forms in the wake of the droplet is leading to a cusp in the flame which is absent in 3-D cases. Such boundary layer development and the flow separation suggest a stronger influence of the wake dynamics on the combustion. Accordingly, there is a stronger interaction between the droplet shape and the combustion dynamics in 2-D when compared to 3-D. 

\begin{figure}[h!]
   \centering
   \includegraphics[width=1\linewidth]{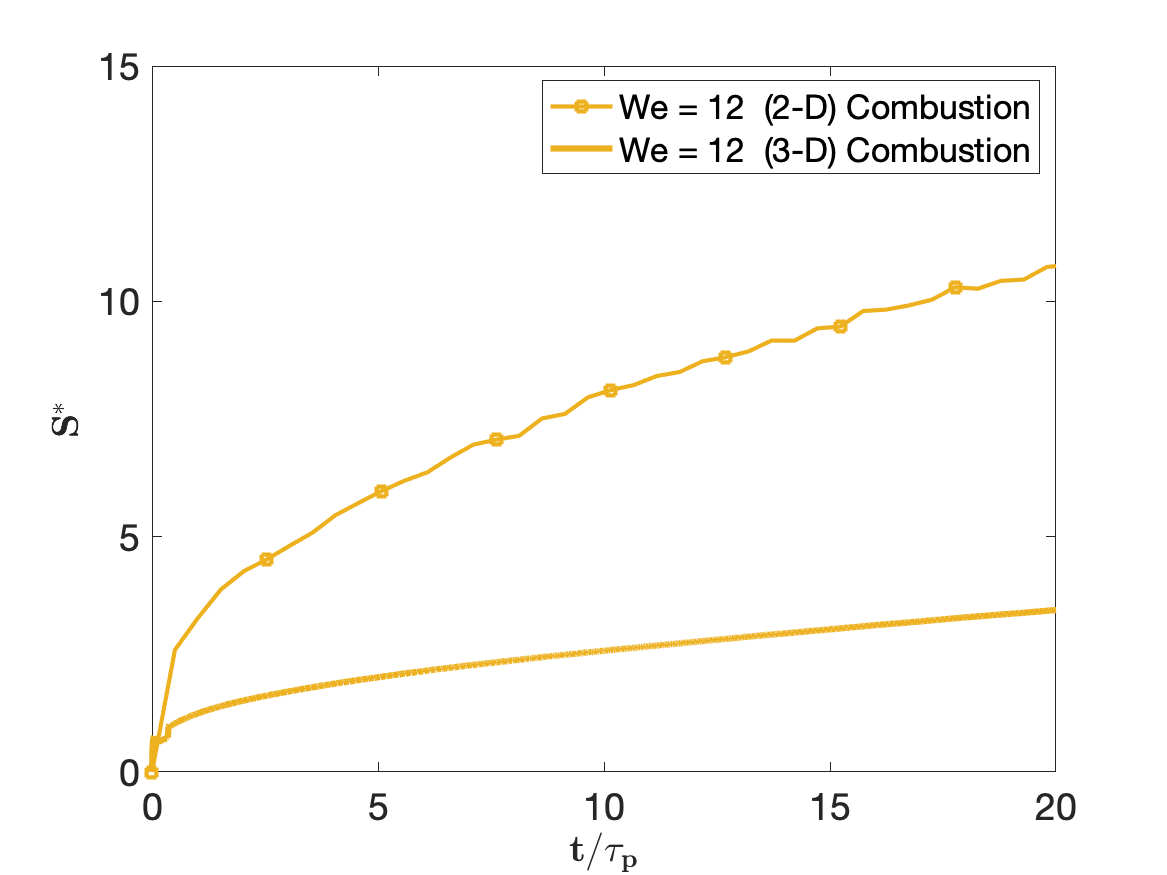} 
   \caption{2-D vs 3-D: Normalized flame stand-off distance ($S^*$) vs $t/\tau_p$ at $Re_0=25$ }
   \label{fig:2D_3D_flame_Lc}
\end{figure}

It is also interesting to note that, although a cusp was not observed in the 3-D flame \cref{fig:3d_cmbst_vectors} (i.e. the stochiometric contour), it is observed on the inner contour of $CO_2$ and absent on the outer contour of $CO_2$ \cref{fig:3d_cmbst_we_12_gT}, suggesting the nature of the flame development may be different between 2-D and 3-D.
Moreover, the flame appears to be wider in 2-D combustion case. The normalized flame stand-off distance ${S}^*$ is calculated using the following expressions,
\begin{equation}
     ({S}^*)_{2-D} = \frac{A_{flame}}{A_{0_{\ droplet}}}
\end{equation}
\begin{equation}
    {(S}^*)_{3-D} = \sqrt{\frac{A_{flame}}{A_{0_{\ droplet}}}}
\end{equation}
Here $A_{0_{\ droplet}}$ is initial area of droplet. It is important to note that the `area' of the 2-D flame is its the surface area per unit depth i.e. its perimeter. The given expressions are written such that for a spherical (circular) flame in 3-D (2-D) the resulting expression simplifies to $r_{flame}/r_0$.
Using these expressions, the flame stand-off distance history is plotted with non-dimensional time as shown in \cref{fig:2D_3D_flame_Lc}. Higher flame stand-off distance is observed for 2-D cases in comparison to 3-D with combustion. This supports the observation of $34\%$ wider flame in 2-D in \cref{fig:2d_3d_cmbst_vectors}. Moreover, the magnitude of ${S}^*$ is seen to be increasing with time which suggests the flame growth in time.

It is important to note that the results here are demonstrated for a envelope flame configuration. Given the greater importance of wake dynamics in 2-D compared to 3-D, it is likely that wake flame dynamics and flame stability will be significantly different between 2-D and 3-D. However, such aspects are beyond the focus of this study.

\section{Conclusions} \label{sec:conclusion}
With the focus on large spray droplets, this work presents a DNS study on the combustion of a freely deformable droplet in a high pressure ($P_\infty = 20 \ atm$), high temperature ($T_\infty = 1400 \ K$) convective flow at $Re_0 = 25$. This work improves upon the existing understanding of droplet evaporation and combustion as most of the past studies consider the droplet shape to be either spherical or an imposed shape such as ellipsoid throughout its lifetime. However, in reality, these large droplets have tendency to deform due to imbalance of surface tension and aerodynamic forces. This deformation is governed by Weber number. $We_0=1$ (nearly spherical shape) and $We_0=12$ (the limit for droplet breakup) are selected for this work. A single step combustion mechanism is used for combustion.

The effect of droplet shape on pure evaporation and combustion is studied by varying the surface tension inside the liquid droplet at $Re_0=25$ for the same ambient gas conditions. The results showed higher total evaporation rate for higher Weber number case. This enhancement reaches upto $ 9 \%$ for $We_0=12$ combustion case. This is due to net effect of decrease in local flux and increase in total surface area for higher Weber number case. 
Moreover, at this $Re_0$ and ambient gas conditions, initial Damk\"{o}hler number is $Da_0 > 1.02 $ which favors a non-spherical envelope flame. From the detailed analysis of gas temperatures, flame shape, and  flame stand-off distance, it is found that even with the droplet shape change and enhancement in total evaporation rate, the combustion demonstrates little sensitivity to droplet shape at this $Re_0$. 
Small differences are noticed in the flame shape, and in particular the flame appears to be longer in the higher Weber number case. However, due to domination of chemical reaction rate over evaporation at given ambient conditions, the overall effect is small. \textcolor{black}{It is also important to note that droplets do not reach critical deformation conditions and break-up even at further higher Weber numbers such as $We_0= 16,\ 20$. Whereas, the non-combusting, non-evaporating droplets at $We_0=16, \ 20 $ demonstrate larger deformation and experience the vibrational break-up mode. This leads to an understanding that the droplet deformation is suppressed by the higher evaporation rates during combustion. Therefore, there is a more complex interaction between droplet shape and combustion which needs further investigation.} 

3-D numerical studies including the combustion are known to be computationally expensive. The question ``what can be learned from less expensive 2-D simulations ? " was explored.
After a careful scrutiny between 2-D and 3-D results at $We_0=12$, it is found that surface averaged local flux is $42.5 \%$ lower in 2-D when compared to 3-D. This is due to larger temperature gradient near the droplet, which drives higher evaporation. Moreover, the trend of total surface area of droplet in 2-D is not same as 3-D.
Though both cases show envelope flame, the presence of larger recirculation region due to larger deformation is causing a cusp in the flame shape at the downstream in 2-D cases. This wake region is significantly smaller and the vortices in the wake are less powerful in 3-D. This leads to absence of cusp in flame.

Moreover, the flame stand-off distance is higher in 2-D when compared to 3-D. 
Hence, it is learned that 2-D studies do not represent the accurate flow physics. 
Although, similar trends are demonstrated for this envelope flame topology. However, it is likely that due to the stronger influence of wake dynamics in 2-D, this strategy will not work well for other flame regimes.

\bibliographystyle{asmems4}

\begin{acknowledgment}
This work has been partially supported by Institute for Critical Technology and Applied Science at Virginia Tech. The authors acknowledge Advanced Research Computing at Virginia Tech for providing computational resources and technical support that have contributed to the results reported in this article.
\end{acknowledgment}

%
% ...

\bibliography{asme2e}



\end{document}